\documentclass[a4paper,11pt]{article}

\usepackage{jcappub} 

\usepackage{upquote,textcomp}

\usepackage[T1]{fontenc} 

\usepackage{amsmath, amsthm, amssymb}
\usepackage{color}
\usepackage{subfigure}
\usepackage{bm}
\newcommand{\be}{\begin{equation}}
\newcommand{\ee}{\end{equation}}
\newcommand{\ba}{\begin{eqnarray}}
\newcommand{\ea}{\end{eqnarray}}

\newcommand{\n}[1]{\label{#1}}
\newcommand{\eq}[1]{(\ref{#1})}

\usepackage{appendix}
\begin{document}
\title{\begin{huge}Spectral Methods in
General Relativity \\~~~~~~~~~~~~~~~~~~~~~~~and\\
Large Randall-Sundrum II Black Holes\end{huge}}
\author[a,b]{Shohreh Abdolrahimi,}
\emailAdd{abdolrah@ualberta.ca}
\author[a,c]{C\'{e}line Catto\"{e}n,}
\emailAdd{celine.cattoen-gilbert@canterbury.ac.nz}
\author[a]{Don N. Page,}
\emailAdd{dpage@ualberta.ca}
\author[a]{Shima Yaghoobpour-Tari}
\emailAdd{yaghoobp@ualberta.ca}
\emailAdd{~~~~~~~~~~~~~~~~~~~~~~~~~~~
~~~~~~~~~~~~~~~~~~~~~~~~~~~~~~~~~~~~~~~~~~~~~~~~~~~~~~~~~~~~~~~~~~Alberta Thy 14-12 }
\affiliation[a]{Department of Physics, 4-181 CCIS, University of Alberta,
Edmonton, Alberta T6G 2E1, Canada}
\affiliation[b]{Institut f\"ur Physik, Universit\"at Oldenburg, Postfach 2503 D-26111 Oldenburg, 
Germany}
\affiliation[c]{BlueFern Supercomputing Unit, University of Canterbury, Christchurch 8140, New Zealand}
\keywords{gravity, modified gravity, and astrophysical black holes\\ \\
{\bf Date:}  Mayan Long Count Calendar 13.0.0.0.0, or day 1,872,000, end of the 13th b\textquotesingle ak\textquotesingle tun (2012 Dec.\ 21 in the Gregorian Calendar) }
 
\abstract{Using a novel numerical spectral method, we have found solutions for large static Randall-Sundrum II (RSII) black holes by perturbing a numerical AdS$_5$-CFT$_4$ solution to the Einstein equation with a negative cosmological constant $\Lambda$ that is asymptotically conformal to the Schwarzschild metric.  We used a numerical spectral method independent of the Ricci-DeTurck-flow method used by Figueras, Lucietti, and Wiseman for a similar numerical solution.  We have compared our black-hole solution to the one Figueras and Wiseman have derived by perturbing their numerical AdS$_5$-CFT$_4$ solution, showing that our solution agrees closely with theirs.  We have also deduced the new results that to first order in $1/(-\Lambda M^2)$, the Hawking temperature and entropy of an RSII static black hole have the same values as the Schwarzschild metric with the same mass, but the horizon area is increased by about $4.7/(-\Lambda)$.}

\maketitle
\flushbottom
\section{Introduction}\label{sec1}

In this paper we consider the Randall-Sundrum II (RSII) \cite{B} model, which is a five-dimensional gravitational model.
The idea of extra dimensions goes back to 1914, when Nordstr\"om made an unsuccessful attempt to describe gravity and electromagnetism
simultaneously by introducing one extra spatial dimension \cite{Nor}. His
theory did not turn out to be correct and was replaced by Einstein's theory. Later, Kaluza's work gave birth to the modern
Kaluza-Klein (KK) theories \cite{Ka, Kl}. Kaluza considered a five-dimensional spacetime with one additional spatial dimension to unify the fundamental forces of gravity and electromagnetism. Initially these theories were rather a mathematical
exercise. The formulation of string theory and M-theory in a space-time with a
number of dimensions greater than four has provided more support for the idea of
higher dimensions. At first, the size of these extra dimensions was naturally considered to be of the order of Planck length, $l_{pl}\sim 10^{-33}\text{ cm}$. This is because in KK theories, using extra compact dimensions, a tower of four-dimensional particles with masses proportional to the inverse size of the compact dimension $L^{-1}$ are produced. However, the standard model has been successfully tested up to $\sim 100 ~\text{GeV}$. In 1983 Rubakov and Shaposhnikov \cite{Ru} proposed a novel model in which fermions and bosons are confined to a four-dimensional subspace of a higher dimensional space-time. Following a similar direction, D-branes have been introduced in string theory \cite{Po}, where fermions, bosons and gauge fields associated with open strings are confined to propagate only along the brane, while gravity, associated with closed strings, can propagate in the bulk. This gave rise to the so-called braneworld models. In simple words a braneworld is a slice through spacetime on which we live and where our standard model physics is confined. Braneworld scenarios also provide a geometrical interpretation of the hierarchy problem, where the electroweak scale $m_{EW} \sim 1~~\text{TeV}$ is much smaller than the Planck scale $M_{Pl}=G^{-1/2}= 1.2\times 10^{16} ~~\text{TeV}$. Two of the most popular braneworld models are those of Randall and Sundrum \cite{A, B}.

In the Randall-Sundrum (RS) models our world is considered as a brane or a domain wall embedded in a five-dimensional spacetime with a negative cosmological constant. The Randall-Sandrum model II (RSII) \cite{B} has one domain wall, situated in a five-dimensional bulk asymptotically anti-de Sitter (AdS) space-time. The ground state bulk metric is precisely AdS on each side of the brane at $w=0$,
\be
ds^2=\frac{l^2}{(l+|w|)^{2}}[dw^2-dt^2+dx_1^2+dx_2^2+dx_3^2],\n{1.1}
\ee
where $l$ is the curvature length scale of the negatively curved five-dimensional AdS space-time, related to the bulk cosmological constant by $l^2=-{6}/{\Lambda}$. This bulk metric satisfies the five-dimensional Einstein equations with a cosmological constant
\be
G_{\mu\nu}=-\Lambda g_{\mu\nu}, ~~~~\text{or} ~~~
R_{\mu\nu}=\frac{2}{3}\Lambda g_{\mu\nu}.\n{1.2}
\ee
Each slice of $w=\text{const}.$ represents a flat four-dimensional space-time. The  comformal factor, which depends on $w$ alone in this case, is known as the warp factor. At $w=0$ we have the flat Minkowski metric of  a domain wall, or brane. This brane is an infinitely thin brane, which satisfies the Israel junction condition
\be
[K_{\mu\nu}]=-\frac{2}{l}\gamma_{\mu\nu},\n{1.3}
\ee
where $[K_{\mu\nu}]$ is the difference of the extrinsic curvature of the brane on the two sides ($w<0$ and $w>0$) and $\gamma_{\mu\nu}$ is the induced metric on the brane at $w=0$. The brane tension is $\sigma={3}/{(4\pi l G_5)}$, where $G_5$ is the five-dimensional Newton's constant. The bulk Einstein equation plus the Israel junction condition for a brane with the RSII value of the tension and without matter imply that the Ricci scalar of the brane metric is zero for a general asymmetric static metric. Applying linear perturbation theory for the metric \eq{1.1} shows that a Newtonian potential can be reproduced, so four-dimensional gravity is recovered for this theory. 

One important question is whether the RSII model can describe our universe. For the RSII model to be a candidate for describing our universe, it must admit black hole solutions. There has been much debate, and various conjectures and claims in the last few years, about the existence of a static black hole solution within the RSII model. An exact black hole solution was found 
for the dimension $d=4$ of the bulk spacetime, where the static black hole is localized on a 2-brane \cite{Emp}, using the so-called C metric \cite{Cm}. In  \cite{Emp2, Tan} it was conjectured that based on the AdS/CFT correspondence large static five-dimensional black hole cannot exist. According to \cite{Emp2}, such bulk black holes
would necessarily be time dependent, since their duals
would describe quantum corrected black holes in a $d-1$
dimensional braneworld. However, counterarguments were
given in \cite{Fit}. In \cite{Ku1,Ku2} nontrivial localized black hole solutions have been found numerically, whose
horizon radii were small compared to the bulk curvature
scale. In their formulation, the problem was reduced to elliptic equations for metric
functions with appropriate boundary conditions. They solved this problem by a relaxation
method. Although their method has worked well for
the small localized black holes, they could not succeed in
finding black hole solutions with large horizon radius. In fact as the mass of the black
hole became large, the convergence became worse and the error grew. One can consider the results \cite{Ku1, Ku2} as evidence for the existence of solutions of a black hole on a brane with a small mass. However, the other interpretation suggested in \cite {Yosh} was that maybe that the growth of the error can be regarded as evidence for the nonexistence of
such solutions. Therefore \cite{Yosh} took the task of re-examining of the result of \cite{Ku1, Ku2}, numerically developing a code having an almost 4th-order accuracy. According to the author, this code derived a highly accurate result for the case where
the brane tension was zero, i.e., the spherically symmetric case. However, a nonsystematic
error was detected in the cases where the brane tension was nonzero. This error was irremovable
by any systematic methods such as increasing the resolution, setting the outer boundary at
more distant location, or improving the convergence of the numerical relaxation. Thus, it was suggested in \cite{Yosh} that a solution sequence of a static black hole on an asymptotically flat
brane that is reduced to the Schwarzschild black hole in the
zero tension limit is unlikely to exist. In  \cite{KL} the result of \cite{Ku1,  Ku2, Yosh} was re-examined again employing a different numerical methods, and the result of \cite{Yosh} was confirmed. For other discussions see \cite{1, 2, 3, 4, 5, 6}. 

Despite all these claims, Figueras and Wiseman \cite{FW} (henceforth FW) recently found such solutions by perturbing an AdS$_5$-CFT$_4$ solution that Figueras, Lucietti, and Wiseman \cite{FLW} (henceforth FLW) had found earlier by Ricci-DeTurck flow.  This AdS$_5$-CFT$_4$ metric is a solution to the Einstein equations with a negative cosmological constant $\Lambda$ that is asymptotically conformal to the Schwarzschild metric.  Because the Schwarzschild metric appears at an AdS$_5$ boundary with an infinite scale factor, it may be viewed as a black hole of infinite mass.

We had independently searched for and found the infinite-mass black hole solution by a different numerical method and were preparing to perturb it to get large-mass RSII black hole solutions when the Figueras et al.\ papers appeared. Here expanding upon \cite{our}, we report that our numerical solution agrees well with that of Figueras et al.\ and thus helps to support the existence of large RSII black holes, despite the doubts expressed by previous work. We point out here that the previous results of \cite{Ku1, Ku2, Yosh, KL} have been concentrated on the small black hole regime. 

We used a spectral method, expressing the components of the 5-dimensional metric in terms of Legendre polynomials in the two nontrivial coordinates, with the appropriate boundary conditions imposed. We chose the 210 coefficients of the polynomials to minimize the integrated square of the error of the Einstein equation. We have reduced the integrated square of the error of the Einstein equation by eight orders of magnitude from the case with no free parameters (constant polynomials). This strongly suggests that we are numerically near an exact solution, though of course our limited computational resources meant that we could not use an infinite number of parameters to reduce the numerical error all the way to zero. The integrated square error is based on the Gauss-Legendre quadrature numerical method, and the minimization procedure uses the simplex search method for multivariable functions. This approach to solving the Einstein equations is novel, and the good agreement of our results with the Figueras et al. results illustrates the success of the method, especially in comparison with the failure of various previous numerical attempts.

We present an explicit approximate metric for the black hole on the brane. Using this approximate metric we demonstrate that the area of an RSII black hole on the brane is slightly greater than a black hole in pure four-dimensional general relativity, and to leading order, the relations between the mass, Hawking temperature, and Bekenstein-Hawking entropy are precisely the same as in four-dimensional general relativity. 

\section{Infinite Black Hole Metric}
The first attempt for finding a black hole solution in the RS model was that of Chamblin, Hawking and Reall \cite{CHR}. They have replaced the Minkowski metric in \eq{1.1} with the Schwarzschild metric; one can in fact replace it with any 4-dimensional Ricci-flat metric. The result is 
\be
ds^2=\frac{l^2}{w^2}[dw^2-U(r)dt^2+U(r)^{-1}dr^2+r^2d\Omega^2],\n{1.4}
\ee
where $U(r) = 1 - 2M/r$ and where $d\Omega^2 = d\theta^2 + \sin^2{\theta}d\phi^2$ is the unit two-sphere metric.  Letting $r = 2M/y$ and $w = 2M/v$ and setting $l=1$ gives
\begin{eqnarray}
ds^2=\frac{dv^2}{v^2}+\frac{v^2 dy^2}{y^4(1-y)}-4v^2(1-y)dt^2 +\frac{v^2}{y^2}d\Omega^2.
\end{eqnarray}
The hypersurfaces of constant $v$ are Schwarzschild metrics of mass $m(v) = v/2$.  The curvature at $y > 0$ diverges at $v=0$, so this black string metric is singular.  We modify the metric by adding some $y^2$ terms to remove this singularity, and we also introduce four metric functions to give
\begin{eqnarray}
ds^2 = &&A\frac{dv^2}{v^2 + y^2} +B\frac{(v^2+y^2)dy^2}{y^4(1-y)}-4C(v^2+y^2)(1-y)dt^2 +D\frac{v^2}{y^2}d\Omega ^2.
\end{eqnarray}
We then replace $v$, which ranges from 0 to $\infty$, by $x = y^2/(y^2 + v^2)$, so that the metric becomes
\begin{eqnarray}
ds^2\!=\!A(1-x){\left[\frac{dx}{2x(1-x)}-\frac{dy}{y}\right]}^2
         \!+\! B\frac{dy^2}{xy^2(1-y)} \!-\! 4C\frac{y^2(1-y)}{x}dt^2\!+\! D\frac{1-x}{x}d\Omega ^2,\n{m.1}
\end{eqnarray}
where $0\leq x\leq1$, $0\leq y\leq1$ and $A(x,y)$, $B(x,y)$, $C(x,y)$ and $D(x,y)$ are smooth functions of $x$ and $y$.  The coordinate boundaries are these:  $x=0$ is the infinite AdS boundary that is conformal to the Schwarzshild metric when we impose $A = B = C = D = 1$ there, $y=0$ is the extremal Poincare horizon, $x=1$ is the axis of symmetry where the two-sphere shrinks to zero size and where we impose $A = D$ for regularity, and $y=1$ is the black hole horizon where we impose the regularity requirement $B = C$.

In the rest of this paper we choose units in which $l=1$, or $\Lambda=-6$.
We impose these regularity conditions and also solve the Einstein equations to lowest order in $x$ by writing $A$, $B$, $C$, and $D$ in terms of polynomial functions of $x$ and/or $y$, $f(y)$, $g(y)$, $\tilde{A}(x,y)$, $\tilde{B}(x,y)$, $\tilde{C}(x,y)$, and $\tilde{D}(x,y)$ with the following forms:
\begin{eqnarray}
A\!&=&\!1\!-\!x(1\!-\!x)(1\!+\!2f)\!+\!x^2g\!+\!x^2(1\!-\!x)\tilde{A},
\nonumber \\ 
B\!&=&\!1\!+\!x f\!+\!x^2\tilde{B}, \nonumber \\
C\!&=&\!1\!+\!x f\!+\!x^2\tilde{B}\!+\!x^2(1\!-\!y)\tilde{C},
\nonumber \\
D\!&=&\!1\!+\!x(1\!-\!x)(1\!+\!f)\!+\!x^2g\!+\!x^2(1\!-\!x)\tilde{D}.\n{8m}
\end{eqnarray}

\begin{table}
\begin{center}
\begin{tabular}{|c|c|c|}
\hline
~~~ & coefficients & I \\\hline
$A=B=C=1$ & 0  &$ 4038$ \\ \hline
0th-order &6 & $69.6986$\\\hline
1st-order & 20& $1.5656$\\\hline
2nd-order & 42& $4.7786\times 10^{-1}$\\\hline
3th-order & 72&$2.3903\times 10^{-2}$\\\hline
4th-order  & 110 & $2.5352\times 10^{-3}$\\\hline
5th-order & 156&$5.4220\times 10^{-4}$\\\hline
6th-order & 210& $4.2385\times 10^{-4}$\\\hline 
\end{tabular}
\end{center}
\caption{The value of $I$, Eq. \eqref{1.8}, vs. the change in the order of the polynomial expansion for the functions $\tilde{A}$, $\tilde{B}$, $\tilde{C}$, $f$, and $g$. Changing the order of the polynomial the number of coefficients which need to be modified change according to the second column.}
\label{TableI}
\end{table}

\setcounter{topnumber}{2}
\setcounter{bottomnumber}{2}
\setcounter{totalnumber}{4}
\renewcommand{\topfraction}{0.85}
\renewcommand{\bottomfraction}{0.85}
\renewcommand{\textfraction}{0.15}
\renewcommand{\floatpagefraction}{0.7}
\begin{figure}
\centering
\hspace{-0.2cm}
\subfigure[$A=B=C=D=1$]{\includegraphics[width=0.45\textwidth]{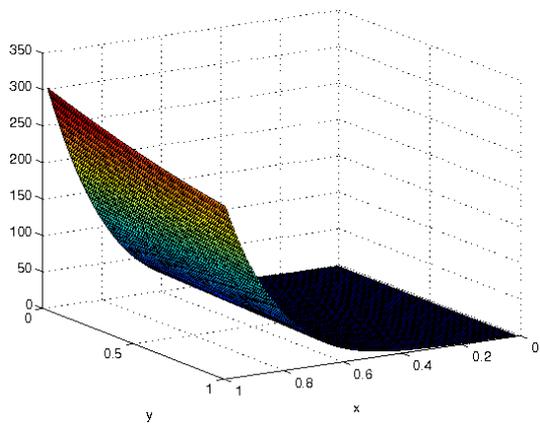}}~~~~~~~~~
\subfigure[6th-order]{\includegraphics[width=0.45\textwidth]{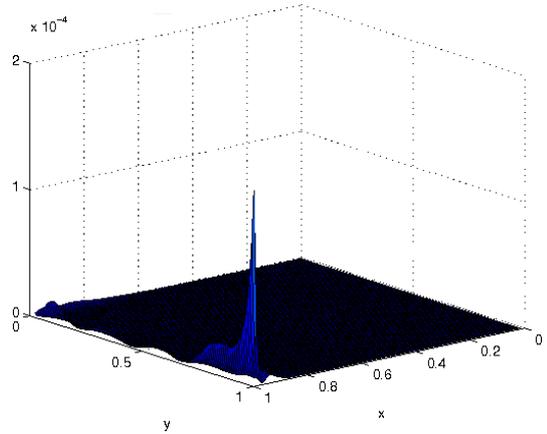}}
\caption{$E_{\mu\nu}E^{\mu\nu}$ vs. $x$ and $y$}
\label{F1.1}
\end{figure}
\begin{figure}
\centering
\subfigure[$A(x,y)$]{\includegraphics[width=0.3\textwidth]{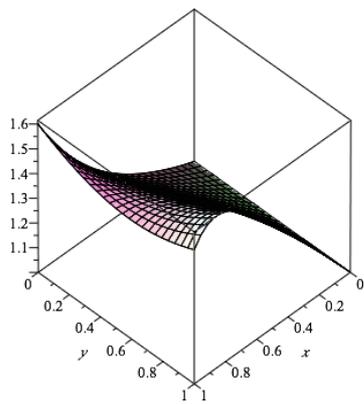}}~~~~~~~~~~~      
\subfigure[$B(x,y)$]{\includegraphics[width=0.3\textwidth]{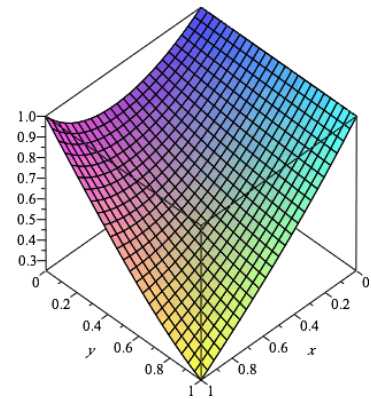}}
\subfigure[$C(x,y)$]{\includegraphics[width=0.3\textwidth]{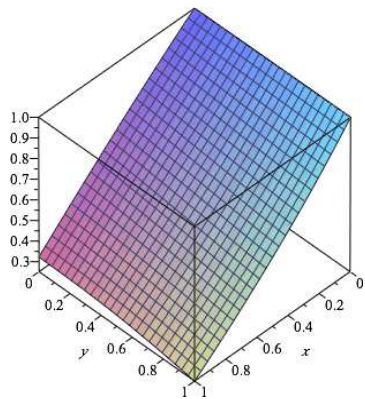}}~~~~~~~~~~
\subfigure[$D(x,y)$]{\includegraphics[width=0.3\textwidth]{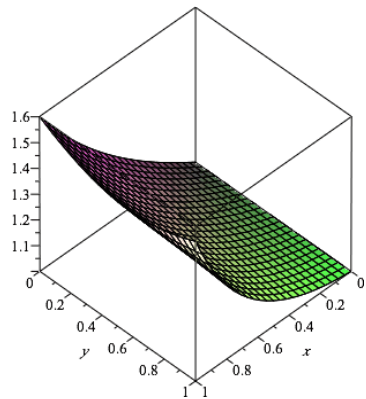}}
\caption{The functions $A$, $B$, $C$, and $D$ vs. $x$ and $y$.}
\label{F8}
\end{figure}
\begin{figure}
\centering
\subfigure[$g_{xx}$]{\includegraphics[width=0.3\textwidth]{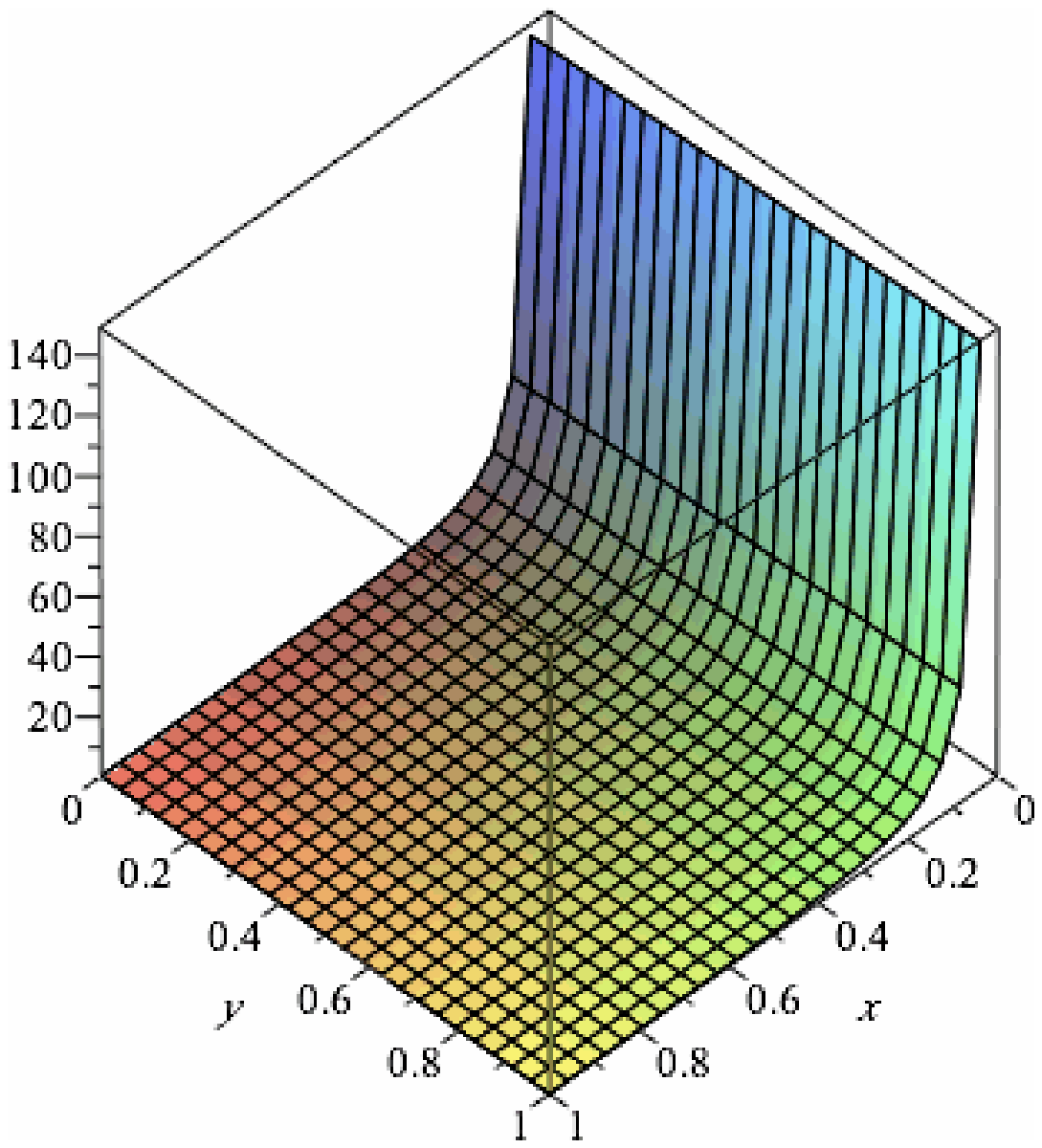}}~~     
\subfigure[$g_{yy}$]{\includegraphics[width=0.3\textwidth]{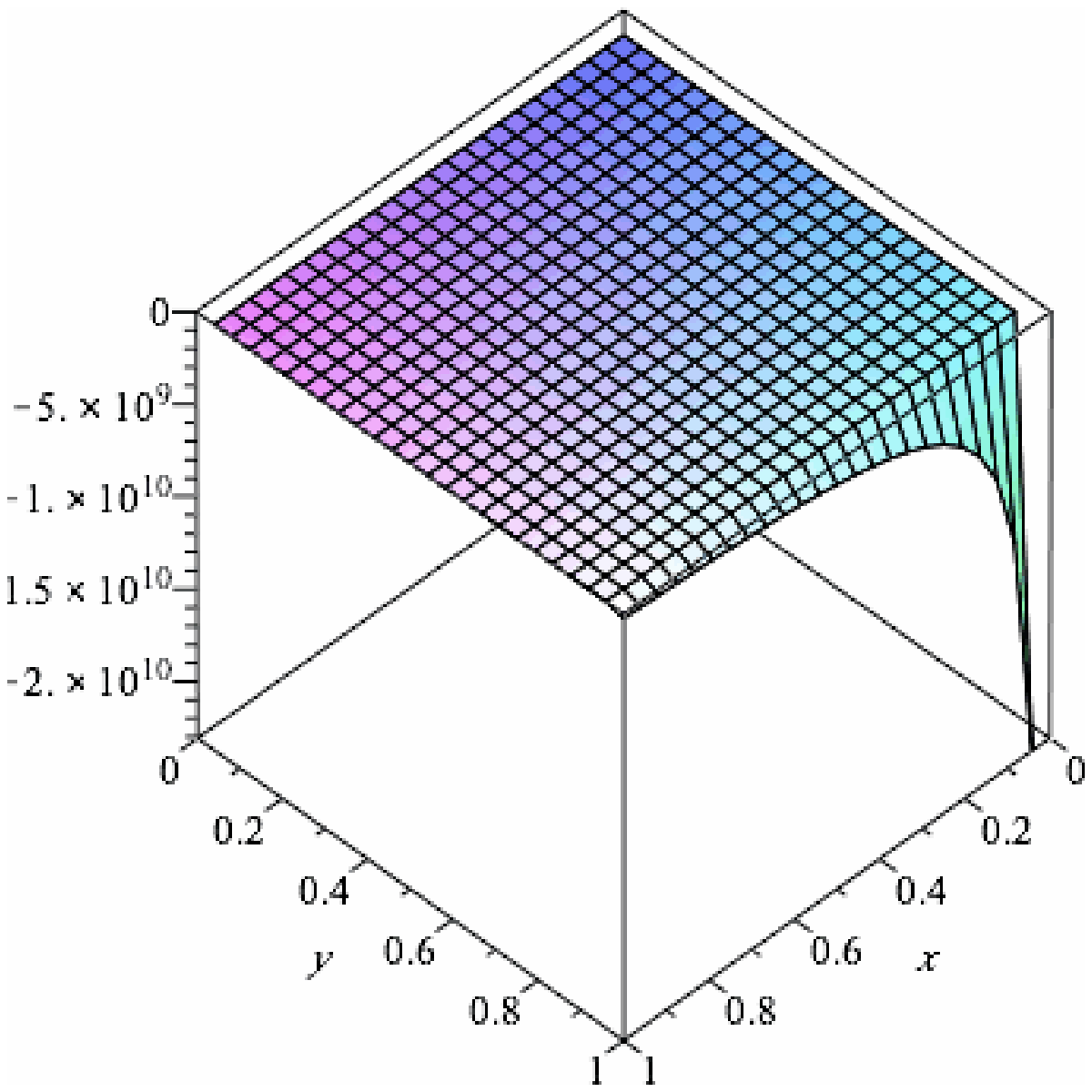}}~
\subfigure[$g_{xy}$]{\includegraphics[width=0.3\textwidth]{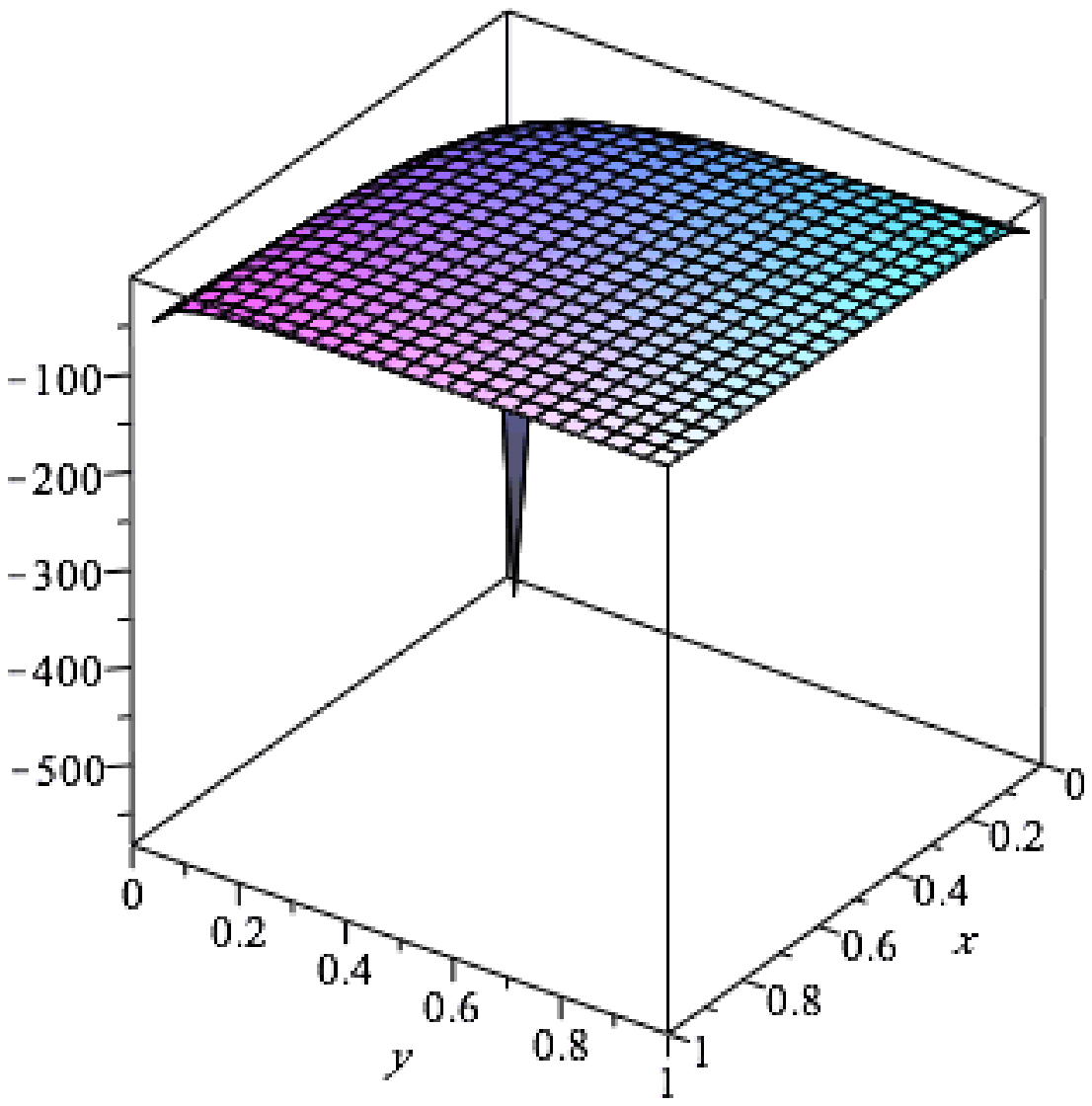}}
\subfigure[$g_{tt}$]{\includegraphics[width=0.3\textwidth]{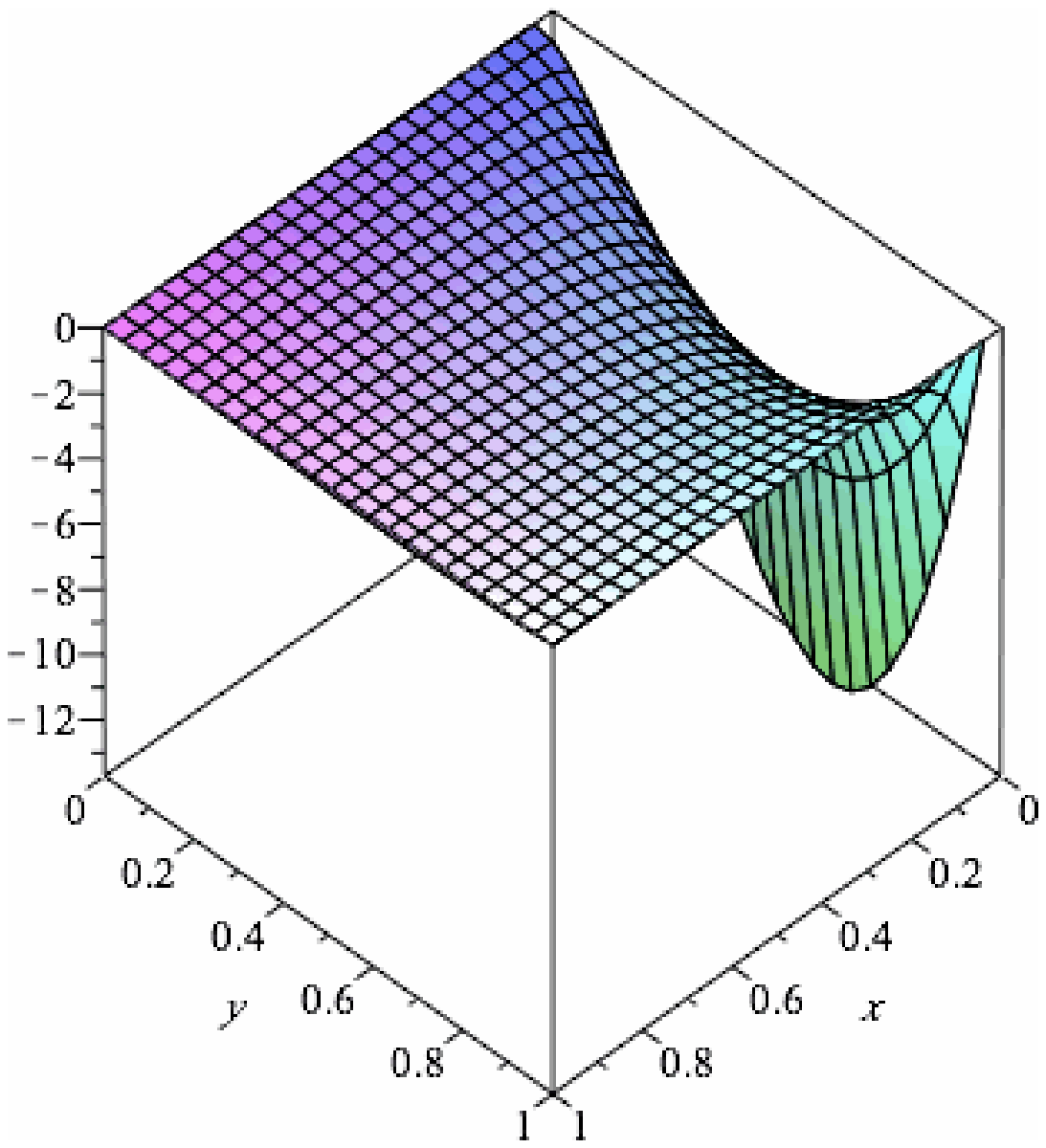}}~~~~~~~~~~
\subfigure[$g_{\theta\theta}$]{\includegraphics[width=0.3\textwidth]{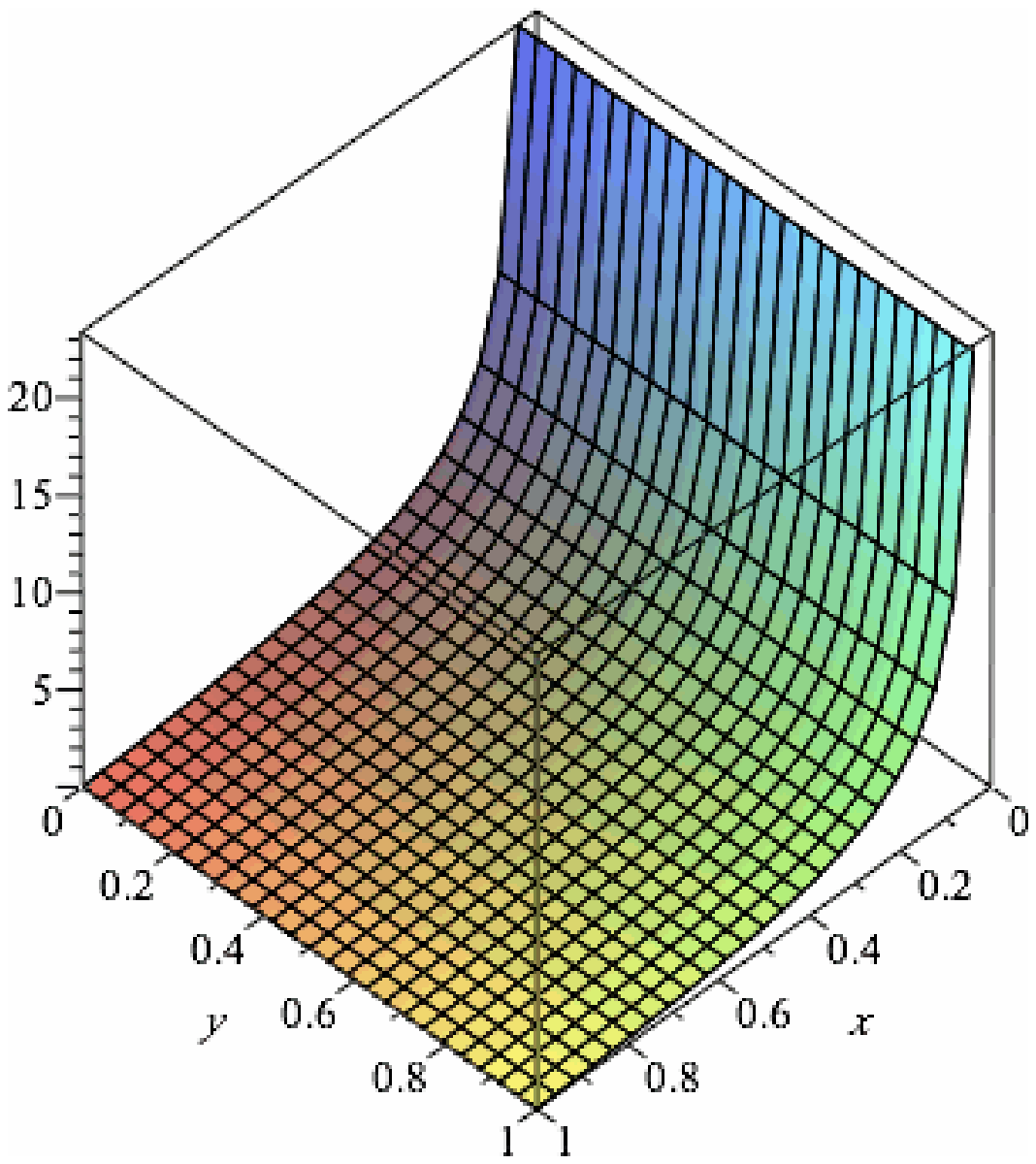}}
\label{F2}
\caption{Metric functions.}
\end{figure}
\begin{figure}
\hspace{-0.2cm}
\subfigure[]{\includegraphics[width=0.45\textwidth]{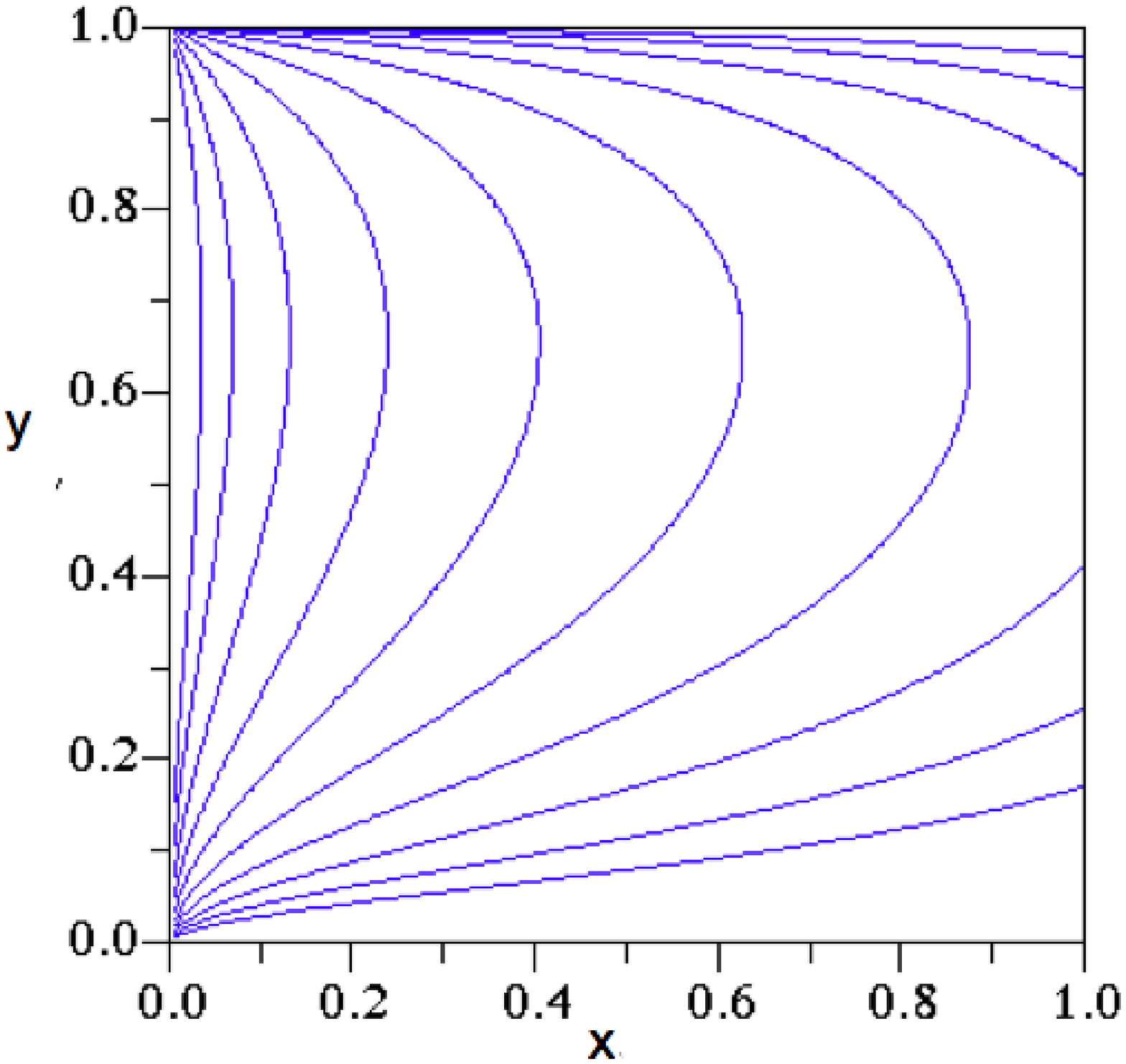}}~~~~~~~~~
\subfigure[]{\includegraphics[width=0.45\textwidth]{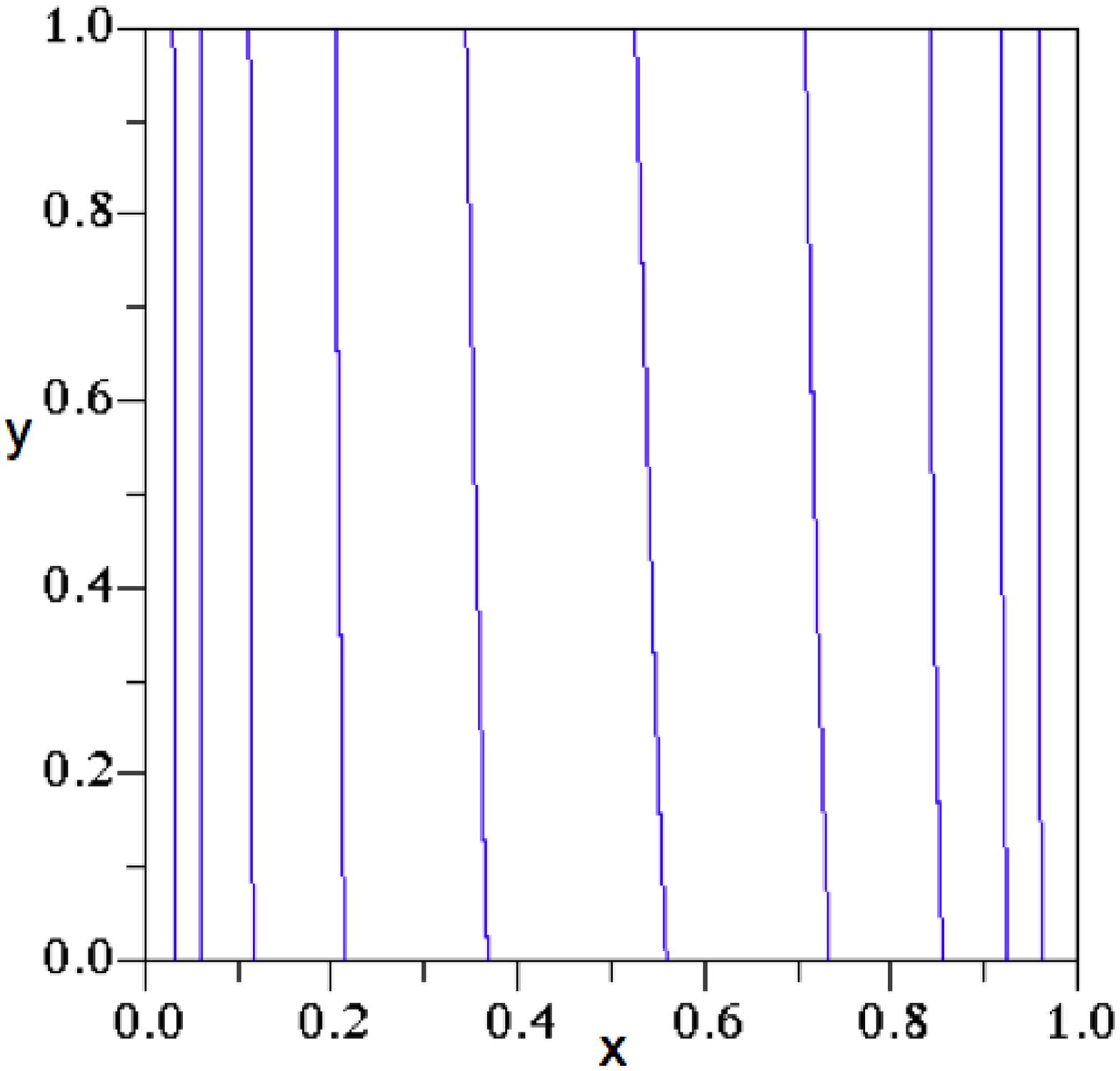}}
\label{Fc1}
\caption{(a) Contour lines of constant $g_{tt}$. The curves from left to right have $g_{tt}$ going from 16 on the left down to $1/32$ on the right with each curve having half the value of the one to its left. (b) Contour lines of constant $g_{\theta\theta}$. The curves from left to right have $g_{\theta\theta}$ going from 32 on the left down to 1/16 on the right with each curve having half the value of the one to its left. }
\end{figure}
\begin{figure}
\begin{center}
\includegraphics[width=0.6\textwidth]{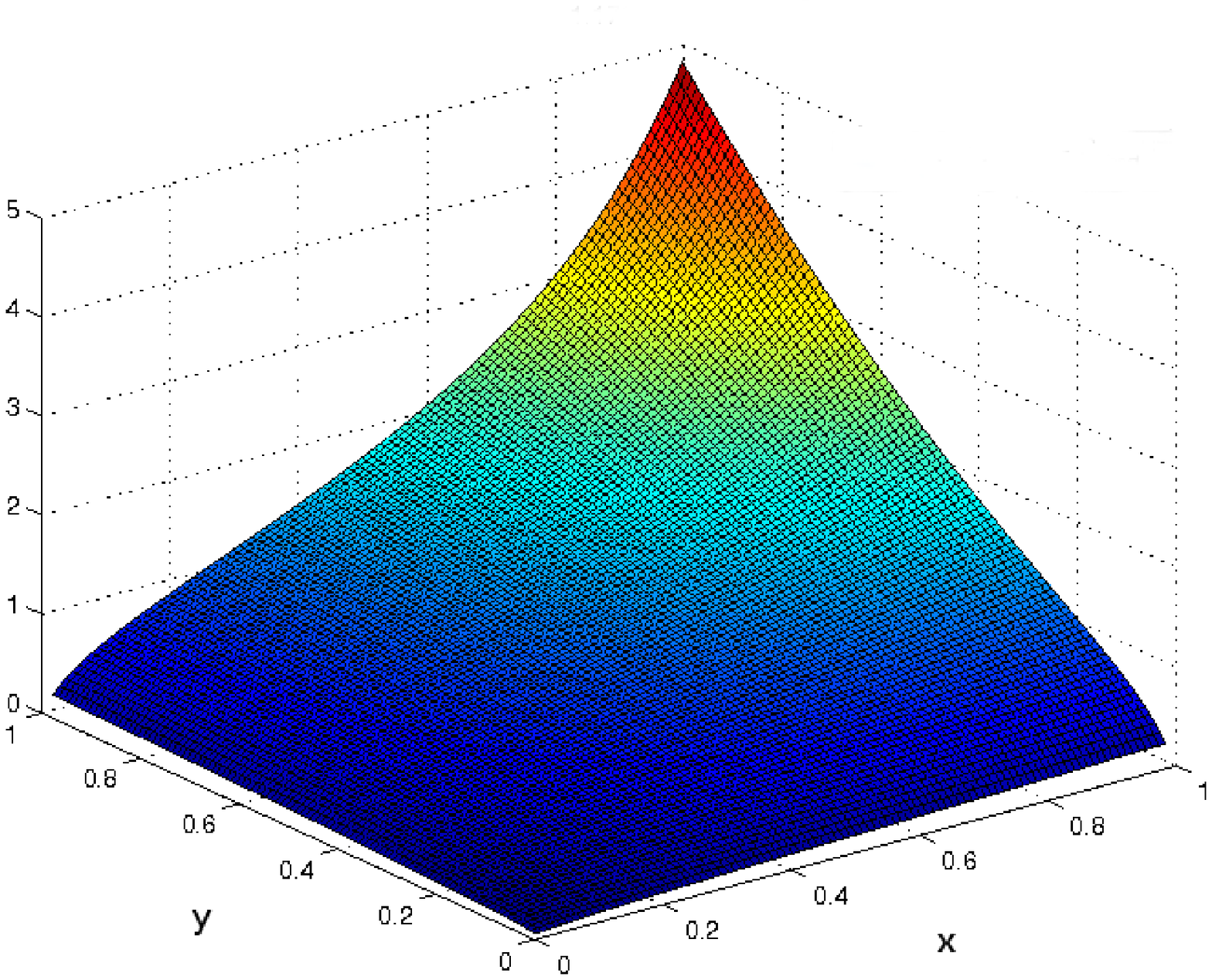}\label{Fw}
\caption{Fourth root of the trace of the square of the Weyl tensor $\mathcal{C}^{1/4}$ as a function of $x$ and $y$. }
\end{center}
\end{figure}
 Our goal is to find the six functions $f,~g,~\tilde{A},~\tilde{B},~\tilde{C}$, and $\tilde{D}$ such that \eq{m.1} solves the Einstein equation \eq{1.2}.  To achieve this numerically, we use  an optimization method. We first define 
 \be
 E_{\mu\nu}=R_{\mu\nu}+4g_{\mu\nu},
 \n{1.7}
 \ee
which should be zero for a solution to the Einstein equation.
Then we define the integrated square error as
 \be
 I=\int\sqrt{-~^{(5)}g} ~E_{\mu\nu}E^{\mu\nu}d^5x, \n{1.8}
 \ee
where this integral is taken over $0\leq x\leq 1$, $0\leq y\leq 1$, $0\leq\theta \leq \pi$, $0\leq\phi\leq  2\pi$, and where we choose $\Delta t = 2\pi$, i.e. the period of an imaginary time coordinate $\tau = it$ avoiding a conical singularity at the horizon location $y=1$. Assuming that $E_{\alpha\beta} E^{\alpha\beta}$ falls off fast enough toward the infinite AdS boundary at $x=0$, where the metric determinant $^{(5)}g \propto 1/x^6$ diverges, the integral converges.  Note that for a static metric, the integrand of \eq{1.8} is positive semidefinite, so for a smooth metric, the error integral $I$ will vanish only if one has a solution to the Einstein equation.

We assume  the functions $f,~g,~\tilde{A},~\tilde{B},~\tilde{C}$, and $\tilde{D}$ to be polynomial functions of $x$ and $y$, and then try to minimize $I$. For the integration method, we use the Gauss-Lobatto quadrature in two dimensions, and for the optimization method we use an unconstrained nonlinear optimization.  We have considered a Taylor series expansion, a shifted-Legendre expansion, and a Pade-Legendre expansion of the $f,~g,~\tilde{A},~\tilde{B},~\tilde{C}$, and $\tilde{D}$ functions in $x$ and $y$.
The Taylor expansion gave a slow rate of convergence for the optimization. The shifted-Legendre and Pade-Legendre give comparable rates; however, Pade-Legendre has a more complicated form. Thus, the shifted-Legendre series expansion, which is defined on the interval $[0,1]$, gives the best result among the three. This is what we chose for our computations. We start with the case where $f=g=\tilde{A}=\tilde{B}=\tilde{C}=\tilde{D}=0$, in other words $A=B=C=D=1$. This gives the error integral $I \approx 4038$. Then we increase the order of polynomial expansion gradually. The 0th-order polynomial expansion gave a minimum $I \approx 69.6986$. For each order, we run the optimization procedure a few times until the value of the integral $I$ remains almost a constant. Table \ref{TableI} shows the final values of $I$ with increasing order of polynomials. Using the 6th-order polynomial expansion of $f,~g,~\tilde{A},~\tilde{B},~\tilde{C}$, and $\tilde{D}$, with a total of 210 coefficients, the integrated squared error was reduced to $0.0004238$, nearly eight orders of magnitude smaller than when $A=B=C=D=1$. Our numerical accuracy is $10^{-10}$. The resulting functions $f,~g,~{A},~{B},~{C}$, and ${D}$ are given in Appendix A. The maximum value of the squared error at any point within the 5-dimensional space-time was then $E_{\alpha\beta}E^{\alpha\beta}= 0.000154$, as shown in Fig  \ref{F1.1}. Thus, we appear to have strong evidence that our numerical method is converging toward an exact solution of the infinite black hole metric. Due to a finite amount of resources and time, we have stopped at this order. 
\begin{figure}
\subfigure[]{\includegraphics[width=0.4\textwidth]{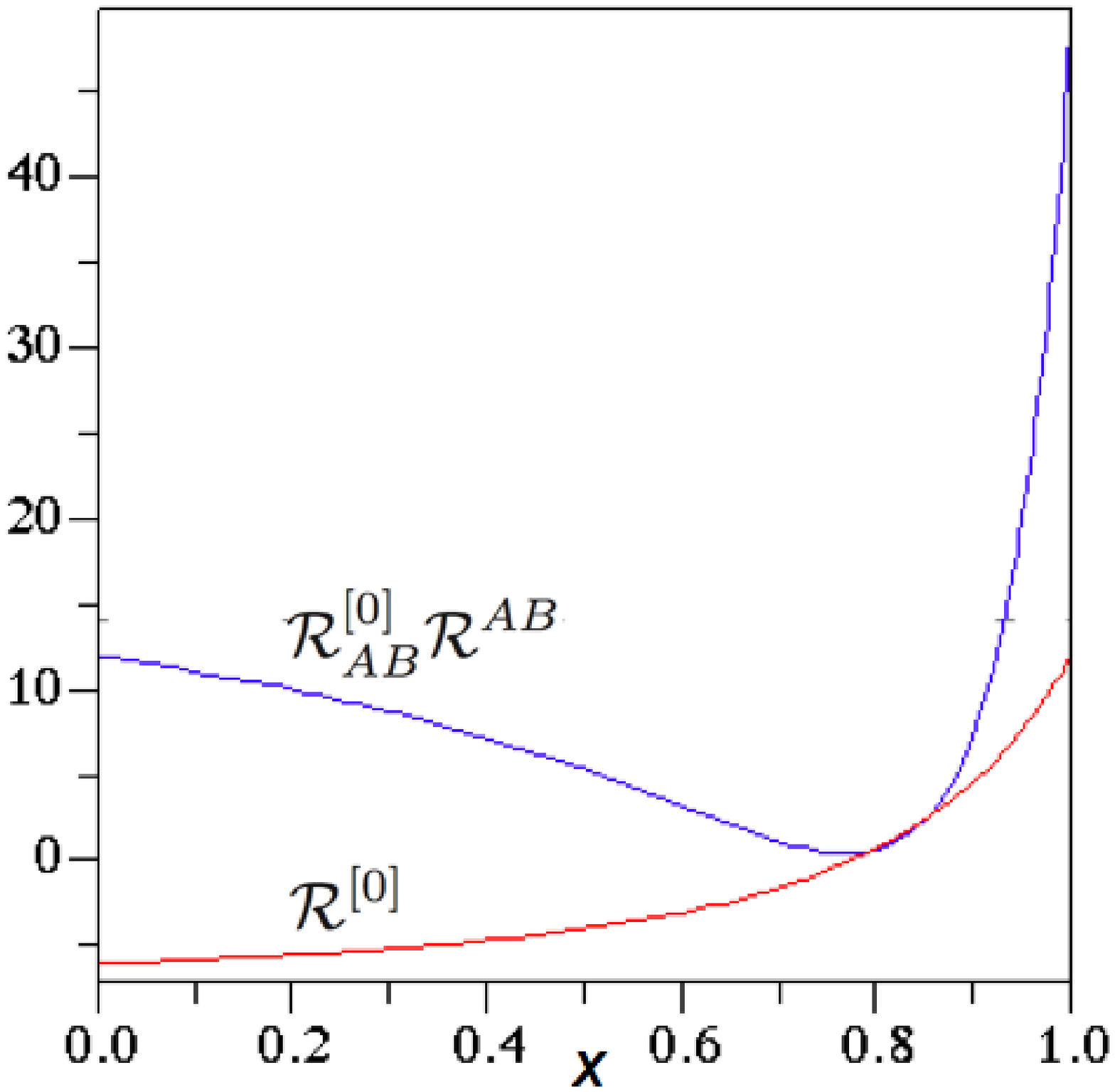}\label{R}}~~~~~~~~~
\subfigure[]{\includegraphics[width=0.4\textwidth]{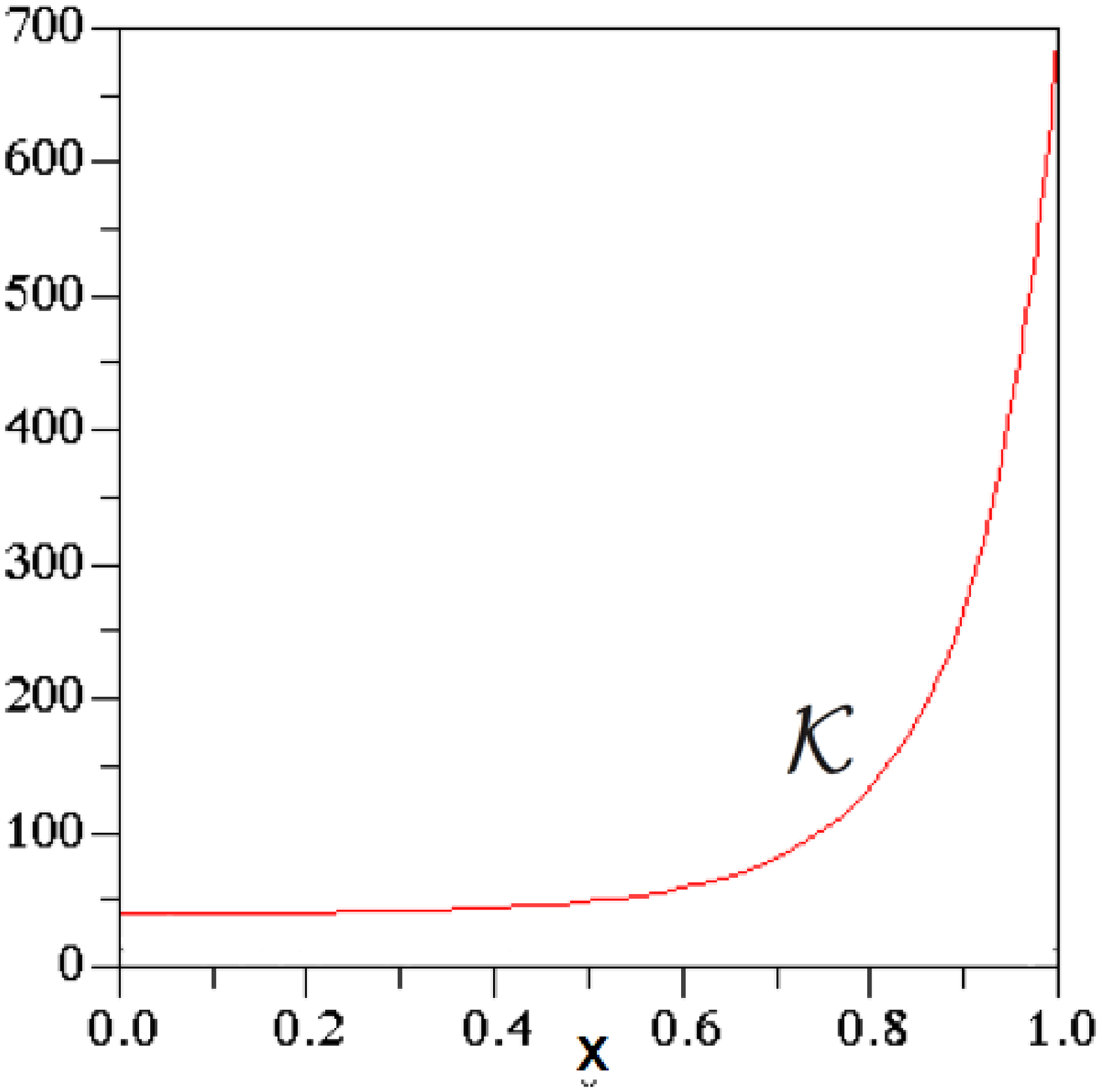}\label{K}}
\caption{(a) Ricci scalar and the square of Ricci tensor of the horizon surface vs. $x$. (b) Space-time Kretschmann scalar calculated on the horizon vs. $x$.}
\label{RK}
\end{figure}
We have calculated the fourth root of the trace of the square of the Weyl tensor, $\mathcal{C}^{1/4}=(C_{\alpha\beta\gamma\delta}C^{\alpha\beta\gamma\delta})^{1/4}$, for our metric \eq{m.1}. Because our metric uses different coordinates from those used by FLW, it is not easy to make many comparisons over the entire bulk five-dimensional manifold.  We have found that the value for the fourth root of the square of the Weyl tensor, $(C_{\alpha\beta\gamma\delta}C^{\alpha\beta\gamma\delta})^{1/4}$ at the corner $x=1$, $y=1$ (the intersection of the black hole horizon with the axis) to be $4.863$ in our metric, which is reasonably near the $5.064$ that FLW privately reported to us from their metric. We shall make many more comparisons below for the four-dimensional large black hole metric.

The metric of a $t=\mathrm{const}$ cross section of the horizon surface at $y=1$ is 
\be
ds_{3}^{2}=g_{AB}dx^{A}dx^{B}
=\frac{A(x)}{4x^{2}(1-x)} dx^{2}+\frac{D(x)(1-x)}{x}d\Omega^{2},
\ee
with
\ba
A(x)&=&1+0.85959x+0.221970x^2-0.66519x^3+0.63763x^4-0.97718x^5\nonumber\\
&+&1.53365x^6-0.96168x^7
-0.29554x^8+0.18927x^9,\\
D(x)&=&1+0.07021x+0.14336x^2+0.22442x^3-0.15951 x^4+0.54809 x^5\nonumber\\
&-&0.71742 x^6+0.60302 x^7-0.19459 x^8+0.02494 x^9.
\ea
We have calculated the Ricci scalar $\mathcal{R}$ and the trace of the square of the Ricci tensor, $\mathcal{R}_{AB}\mathcal{R}^{AB}$, of the horizon surface in Fig \ref{R}. The Ricci scalar at $x=0$ is $-6$ and is equal to $12.17095$ at $x=1$. The square of Ricci tensor is $12$ at $x=0$ and reaches $49.37738$ at $x=1$. 
Using equation (65) of \cite{SA}, which in our case simplifies to 
\be
\mathcal{K}|_{\mathcal{H}}=6\mathcal{R}_{AB}\mathcal{R}^{AB}-4\Lambda \mathcal{R}+\frac{28}{9} \Lambda^{2},
\ee
we give in Fig. \ref{K} the Kretschmann scalar of the 5-dimensional space-time calculated on the horizon.  The Kretschmann scalar at $x=0$ is $40$, and it increases to $700.3672$ at $x=1$. 
\section{Finite Black Hole Metrics}
After doing the optimization, we have the functional form of our metric including the sixth-order polynomials from the numerical calculation. This metric is conformally Schwarzschild with infinite mass at the infinite AdS boundary at $x=0$. In the asymptotically infinite region, the five-dimensional bulk metric near the boundary has the form
\begin{equation} 
ds^2 = dr^2 + e^{2r}\tilde{g}_{\mu\nu}(r,x)dx^\mu dx^\nu,
\end{equation}
where $r$ is the outward proper distance as one approaches the AdS boundary at infinite proper distance, $r \sim \ln{v}$, and $x$ is not the same single coordinate used for the numerical calculation but instead denotes all the other four coordinates besides $r$. Using the Fefferman-Graham (FG) expansion \cite{A1, B1, C1}, $\tilde{g}_{\mu\nu}$ can be written as
\begin{equation}
\label{A}
\tilde{g}_{\mu\nu}(r,x)= g_{\mu\nu}^{(0)}(x)+ e^{-2r}g_{\mu\nu}^{(2)}(x)+ e^{-4r}g_{\mu\nu}^{(4)}(x) - 2e^{-4r} r h_{\mu\nu}^{(4)}(x)+ e^{-4r}t_{\mu\nu}(x) + O(e^{-6r}).
\end{equation}
When the bulk Einstein equation is considered as a second-order differential equation in the radial coordinate $r$, and when $g_{\mu\nu}^{(0)}$ is the conformal metric on the infinite AdS boundary, the boundary conditions are the conformal metric $g_{\mu\nu}^{(0)}$ and a traceless and divergenceless tensor $t_{\mu\nu}(x)$. Then $g_{\mu\nu}^{(2)}=-\frac{1}{2}R_{\mu\nu}^{(0)}(x)+\frac{1}{12}R^{(0)}g_{\mu\nu}^{(0)}(x)$. The terms $g_{\mu\nu}^{(4)}(x)$ and $h_{\mu\nu}^{(4)}(x)$ are functions of the Ricci tensor $R_{\mu\nu}^{(0)}(x)$ of the asymptotic conformal metric $g_{\mu\nu}^{(0)}(x)$; see \cite{C1} for their exact definitions. For our infinite mass black hole metric, we set $g_{\mu\nu}^{(0)}(x) = g_{\mu\nu}^\mathrm{Sch}$ for the RSII brane at $r=\infty$. Therefore, our conformal boundary metric is the Ricci-flat Schwarzschild metric. Then, the only non-zero term in \eqref{A} before the $O(e^{-6r})$ terms is the one including $t_{\mu\nu}(x)$. The metric has the asymptotic form
\begin{equation}
\label{B}
ds^2 = dr^2
 + e^{2r}\left[e^{\beta}\frac{dy^2}{y^4(1-y)}
  -4e^{\gamma}(1-y)dt^2 
  + e^{\delta}\frac{1}{y^2}d\Omega^2\right],
  \end{equation}
\begin{eqnarray}  
\beta &=& e^{-4r} ~{t_r} ^r (y)+ O(e^{-6r}),\\
\gamma &=& e^{-4r} ~{t_t} ^t (y)+ O(e^{-6r}),\\
\delta &=& e^{-4r} ~{t_ \theta} ^\theta(y) + O(e^{-6r}).
\end{eqnarray}
${t_r} ^r$, ${t_t} ^t$, and ${t_ \theta} ^\theta$ are the components of the traceless, divergenceless tensor $t_{\nu}^{\mu}(y)$, which is diagonal in our coordinate system. It can be shown that ${t_r} ^r$, ${t_t} ^t$ and ${t_ \theta} ^\theta$ are the vacuum expectation values of the stress tensor components on the boundary of the CFT energy-momentum tensor, $\langle {{T_r} ^r} \rangle$, $\langle {T_t} ^t \rangle$, and $\langle {T_ \theta} ^\theta \rangle \!=\! \langle {T_\varphi} ^\varphi \rangle$ respectively. We derive approximations for ${t_r} ^r$, ${t_t} ^t$, and ${t_ \theta} ^\theta$ as functions of $y$ from our numerical result for the sixth-order polynomials. For finding the energy-momentum tensors as functions of $y$, we use the coefficients of the first and second orders of $x$ in $A(x,y)$, $B(x,y)$, $C(x,y)$, and $D(x,y)$, which are functions of $y$. We call them $A1(y)$, $B1(y)$, $C1(y)$, and $D1(y)$ for the coefficients of $x$ and $A2(y)$, $B2(y)$, $C2(y)$, and $D2(y)$ for the coefficients of $x^2$. 
Then, we calculate ${t_r} ^r$, ${t_t} ^t$, and ${t_ \theta} ^\theta$ by using these functions with the following formulas:
\begin{eqnarray}
\hspace{-0.8cm}{t_r} ^r(y)&=&-\frac{1}{16}y^4\bigg[11-14y-(10-14y-3A1)A1-4A2-16B2-(12y-13y^2)\frac{dA1}{dy}\nonumber\\
&&-2(y^2-y^3)\frac{d^2A1}{dy^2}\bigg],\label{T1}\\
\hspace{-0.8cm}{t_t} ^t(y)&=&-\frac{1}{16}y^4\bigg[-1-2y+2(1+y+3A1)A1-4A2-16C2+y^2\frac{dA1}{dy}\bigg],\label{T2}\\
\hspace{-0.8cm}{t_ \theta} ^\theta(y)&=&\frac{1}{16}y^4\bigg[5-4y-(6-4y+3A1)A1+4A2-16D1+16D2-2(y-y^2)\frac{dA1}{dy}\bigg].\label{T3}
\end{eqnarray}
Plugging back the functions $A1(y)$, $B1(y)$, $C1(y)$, and $D1(y)$ and $A2(y)$, $B2(y)$, $C2(y)$, and $D2(y)$ from our numerical results into the equations for ${t_r} ^r$, ${t_t} ^t$, and ${t_ \theta} ^\theta$, \eqref{T1}-\eqref{T3} gives us
\begin{eqnarray}
{t_r} ^r(y)&=&-0.01174y^4+0.38148y^5-0.85298y^6+1.36570y^7-2.01560y^8\nonumber\\
&&+1.91160y^9-0.80470y^{10}
+0.08460y^{11}-0.00739y^{12}+0.00496y^{13}\nonumber\\
&&-0.00181y^{14}+0.00023y^{15}-0.00001y^{16},\\
{t_t} ^t(y)&=&+0.00341y^4+0.48856y^5-1.18280y^6+1.10450y^7-0.03021y^8\nonumber\\
&&-0.65406y^9+0.36864y^{10}-0.03968y^{11}-0.00739y^{12}+0.00496y^{13}\nonumber\\
&&-0.00181y^{14}+0.00023y^{15}-0.00001y^{16},\\
{t_ \theta} ^\theta(y)&=& +0.00261y^4-0.40736y^5+0.86236y^6-0.78244y^7+.28004y^8\nonumber\\
&&+0.02035y^9-0.03666y^{10}+0.01027y^{11}-0.00739y^{12}+0.00496y^{13}\nonumber\\
&&-0.00181y^{14}+0.00023y^{15}-0.00001y^{16}.
\end{eqnarray}
The constraints on ${t_r} ^r$, ${t_t} ^t$, and ${t_ \theta} ^\theta$ as the components of a traceless conserved stress-energy tensor are
\begin{equation}
\label{C}
{t_r} ^r+{t_t} ^t +2{t_ \theta} ^\theta=0 ,
\end{equation}
\begin{equation}
\label{D}
2y(1-y )\frac{d}{dy}({t_r} ^r)+y({t_t} ^t-{t_r} ^r)+4(1-y)({t_ \theta} ^\theta-{t_r} ^r)=0 .
\end{equation}
Equation \eqref{C} is the traceless condition for the energy-momentum tensor, $t_\mu^\mu=0$, and equation \eqref{D} corresponds to $\bigtriangledown_\mu t^{\mu \nu}=0$. Checking our numerical results for the energy-momentum conditions, \eqref{C} and \eqref{D}, shows small deviations from zero, which can be explained as numerical error. For the traceless condition, the maximum deviation is $4.57\times 10^{-4}$, and for the $\bigtriangledown_\mu t^{\mu \nu}=0$ condition, it is $1.82\times 10^{-3}$. 
\\\indent Now, we have the infinite mass black hole metric \eqref{B}, numerically. To find the large but finite mass black hole, we need to perturb our infinite mass metric by replacing the RSII brane at the AdS conformal boundary, $r=\infty$, with a brane at finite $r=-\ln~\epsilon$ for $\epsilon=e^{-r}\ll1$. The conformal metric $g_{\mu\nu}^{(0)}(x)$ is perturbed to
\begin{equation}
\label{E}
g_{\mu\nu}^{(0)} = g_{\mu\nu}^\mathrm{Sch} + \epsilon^2 h_{\mu\nu},
\end{equation}
with the perturbation metric $h_{\mu\nu}$. For the metric \eqref{E}, the Ricci tensor is not zero, and the same is true for the $e^{-2r}$ term in the FG expansion. The perturbation would effect the form of $ t^{\mu \nu}(x)$ as well. However, because it is multiplied by $e^{-4r}$ in the FG expansion, and we are working to lowest non-trivial order in $\epsilon^2$, we can still use the values of $ t^{\mu \nu}(x)$ from the infinite mass black hole bulk solution.

The second fundamental form is $K_{\mu \nu}=-\frac{1}{2}\partial_{r}[e^{2r}\tilde{g}_{\mu\nu}(r,x)]$, where we
are using the opposite sign convention from FW, so that
our second fundamental form is half the positive derivative
with respect to distance from the brane of the induced
metric on the four-dimensional hypersurfaces at
each constant distance, with the derivative evaluated at
zero distance from the brane. Therefore, to first order in $\epsilon^2$, the Israel junction condition gives 
\begin{equation}
g_{\mu\nu}^{(2)} = -\frac{1}{2}R_{\mu\nu}^{(0)}
+ \frac{1}{12}R^{(0)}(x)g_{\mu\nu}^{(0)}\\
= -2\epsilon^2 t_{\mu\nu}.
\end{equation} 
As mentioned, up to this order $t^{\mu \nu}$ is traceless. Then the Ricci tensor for the metric \eqref{E} is calculated as
\begin{equation}
\label{F}
R_{\mu\nu}^{(0)} = 4\epsilon^2 t_{\mu\nu}.
\end{equation}
Using equation \eqref{F} with $t_{\mu}^{\nu}(y)$ calculated from the infinite metric leads us to have the $ h_{\mu\nu}$ and spherically symmetric static metric $g_{\mu\nu}^{(0)}$ in \eqref{E}. The induced metric on the brane can be found as
\begin{equation}
\gamma_{\mu\nu} = \frac{1}{\epsilon^2}\tilde{g}_{\mu\nu}
= \frac{1}{\epsilon^2}g_{\mu\nu}^\mathrm{Sch} + h_{\mu\nu}
+ O(\epsilon^2).
\end{equation}
The bulk Einstein equation plus the Israel junction condition on the brane without matter imply that the Ricci scalar for the metric on the brane is zero. To first order in $1/R_0^2 = \epsilon^2 = (3/2)/(-\Lambda M^2)$, the metric on the brane is
\begin{equation}
\small ds^2 =(R_0^2 + 2b)\frac{dy^2}{y^4(1-y)}\\
      - (R_0^2 + 2c)4(1-y)dt^2\\
      + (R_0^2 + 2d)\frac{1}{y^2}d\Omega^2.
\end{equation}
For a general metric on the brane, working in the gauge $h_{t}^{t}=0$, we have
\begin{eqnarray}
&&h_y^y(y) = 2b(y) = -\frac{2y^2(1-y)}{3(4-3y)}\bigg(F+y\frac{dF}{dy}\bigg),\n{hyy}\\
&&h_t^t(y) = 2c(y) =  0,\\
&&h_\theta^\theta(y) = h_\phi^\phi(y) = 2d(y) =  \frac{y^2}{6}F(y),
\label{F36}
\end{eqnarray}
with
\begin{equation}
\label{G}
F(y)  = \frac{2-3y}{(4-3y)^2}
          \int_0^y\frac{(4-3u){t_r} ^r(u)}{u^3(2-3u)^2}du.
\end{equation}
According to \eqref{G}, using ${t_r} ^r$ from the numerical result leads us to find $F(y)$ and the $h_{\mu\nu}$ components afterwards. 

It can be shown that the asymptotic behaviour of $t_{\mu\nu}$ goes as $y^5$, and we know that $t_{\mu\nu}$ is traceless. Considering the above charactristics for $t_{\mu\nu}$, we try to find the traceless conserved fits for ${t_r} ^r$, ${t_t} ^t$, and ${t_ \theta} ^\theta$ back in \eqref{G}. We define $F(y)$ as a polynomial with unknown coefficients, and define ${\hat{t}_r} ^r$, ${\hat{t}_t} ^t$, and ${\hat{t}_ \theta} ^\theta$  as follows:
\begin{eqnarray}
\hspace{-1cm}
{\hat{t}_r} ^r(y)&=&\frac{y^5}{12(4-3y)}\bigg[6(1-y)F+y(2-3y)\frac{dF}{dy}\bigg],\\
\hspace{-1cm}{\hat{t}_ \theta} ^\theta(y)&=& -2{\hat{t}_r} ^r(y)+\hat{\epsilon}(y),\\
\hspace{-1cm}{\hat{t}_t} ^t(y)&=& 3{{\hat{t}_r} ^r}(y)-2\hat{\epsilon}(y),\\
\hspace{-1cm}\hat{\epsilon}(y)&=&\frac{y^5}{12(4-3y)^2}\bigg[12(1-y^2)F-y(12-14y+3y^2)\frac{dF}{dy}-y^2(1-y)(4-3y)\frac{d^2F}{dy^2}\bigg],
\end{eqnarray}
where finding fits for our numerical energy-momentum tensor components as functions of $y$ will help us to compare our results with the FLW results in the next section.
\section{RSII black hole metric on the brane}
We can now write the metric on the brane in terms of the function $F$. 
Define $(2M)^{2}\equiv 1/\epsilon^{2}=6/(-\Lambda \epsilon^{2})$ and the new radial coordinate $\rho=2M/y$. Then, to first order in $\epsilon^{2}=(3/2)/(-\Lambda M^{2})$, the metric on the brane reads
\begin{eqnarray}
^4ds^2 = \gamma_{\mu\nu}dx^\mu dx^\nu 
        &=&\!\! \left[1\! -\! \frac{1}{(\!-\Lambda\rho^2)}
                     \frac{\rho\!-\!2M}{\rho\!-\!1.5M}\!
		     \left(\!F\!-\rho\frac{dF}{d\rho}\!\right)\!\right]
		     \!\left(\!1\! -\! \frac{2M}{\rho}\!\right)^{\!-1}\!d\rho^2 	\nonumber\\	
	&&- \left(1 - \frac{2M}{\rho}\right) dt^2
          + \left[\rho^2 + \frac{1}{(-\Lambda)}F\right]d\Omega^2,\nonumber\\\n{MB}
\end{eqnarray}
where we have rescaled the time coordinate $t$ by a factor of $4M$ so that $\gamma_{tt}=-1$ at $\rho=\infty$.

One can show that to have the leading order of the Einstein equation for the five-dimensional infinite black hole metric \eq{m.1} satisfied at the corner $x=0$ and $\rho\rightarrow \infty$, the asymptotic behavior of $\beta$, $\gamma$, and $\delta$ must be as $1/\rho^{5}$ for $\rho>>2M$. This is in accordance to the result of Figuras and Wiseman \cite{FW} and \cite{FLW}, that $t_{\mu\nu}$ goes as $1/\rho^{5}$. This asymptotic behavior of $t_{\mu\nu}$ implies that $F$ approaches unity as $\rho\rightarrow \infty$. 
To fit to our data $t_{\mu\nu}^{(2)}(x) = t_{\mu\nu}^{\mathrm{our}}(x)$, and to fit to the FLW data, which they have kindly sent us, $t_{\mu\nu}^{(1)}(x) = t_{\mu\nu}^{\mathrm{FLW}}(x)$, we took $F_1 = F_{\mathrm{FLW}}$ and $F_2 = F_{\mathrm{our}}$ to be cubic polynomials in $y \equiv 2M/\rho$ with the constant coefficient set to unity, and $F_{11}$ to be an 11th order polynomial with the constant coefficient set to unity and then chose the other coefficients in each case to minimize the respective
\begin{equation}
J_i \!\!=\frac{\!\! \int\! {\rho^4 \Delta t_{\mu\nu}^{(i)} \Delta t^{\mu\nu}_{(i)} \sqrt{-^{(4)}\gamma} d^4x}}
   { \int\!\! {\rho^4 t_{\mu\nu}^{\mathrm{FLW}} t^{\mu\nu}_{\mathrm{FLW}}      
            \sqrt{-^{(4)}\gamma} d^4x}}.
\n{J}
\end{equation}
\begin{figure}[htb]
\begin{center}
\includegraphics[width=7cm]{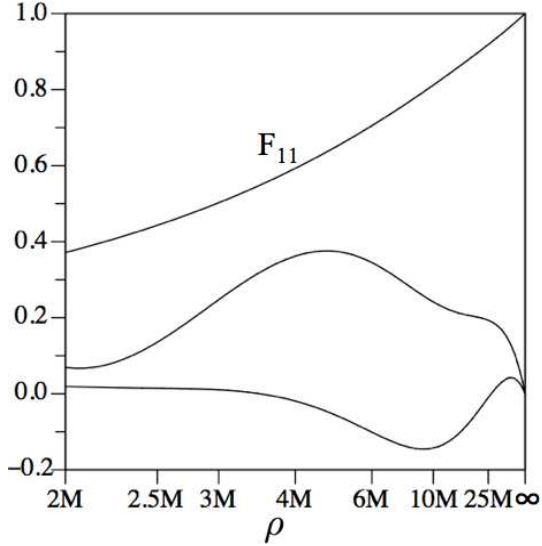}
\caption{At the top is an 11th-order polynomial fit $F_{11}$ to the FLW data, which gave normalized mean-square error $J_{11} = 0.0000572$, 92\% of $J_{\mathrm{FLW}}$.  Because the differences from $F_{11}$ of the cubic fits $F_{\mathrm{FLW}}$ and $F_{\mathrm{our}}$ are too small to show up when plotted directly on this graph, at the bottom we have expanded these differences by a factor of 50 and plotted $50(F_{\mathrm{FLW}} - F_{11})$ (bottom curve) and $50(F_{\mathrm{our}} - F_{11})$ (middle curve).}\label{F4.0} 
\end{center} 
\end{figure} 
For the two values of $i$ ($i=1$ for the FLW data and $i=2$ for our data), in the numerator $\Delta t_{\mu\nu}^{(i)} = t_{\mu\nu}^{F_i} - t_{\mu\nu}^{(i)}$ is the difference between the $t_{\mu\nu}^{F_i}(x)$ given by the cubic for $F_i(x)$ and the $t_{\mu\nu}^{(i)}$ given by the numerical data. To find the 11th order polynomial fit to the FLW data, $\Delta t_{\mu\nu}^{11}= t_{\mu\nu}^{F_{11}} - t_{\mu\nu}^{\mathrm{FLW}}$, the difference between the $t_{\mu\nu}^{F_{11}}(x)$ given by the 11th order fit $F_{11}$ and the $t_{\mu\nu}^{\mathrm{FLW}}$ given by the FLW numerical data. The integral in the denominator was included to make $J$ a normalized mean-square error. It has $t_{\mu\nu}^{\mathrm{FLW}}(x)$ given by the FLW numerical data, the same in each case to give a constant normalizing factor. The factor of $\rho^4$ was included to increase the weight of the large-$\rho$ part, though the integrals are still dominated by the small-$\rho$ part, since $t_{\mu\nu}(x)$ drops off asymptotically as the inverse fifth power of the radial coordinate $\rho$ \cite{FLW,FW}. For small $1/(-\Lambda M^{2})$, Eq. \eq{J} is approximately equivalent to
\begin{equation}
J_i \!\!=\frac{\!\! \int\! {\rho^6 \Delta t_{\mu\nu}^{(i)} \Delta t^{\mu\nu}_{(i)}  d\rho}}
   { \int\!\! {\rho^6 t_{\mu\nu}^{\mathrm{FLW}} t^{\mu\nu}_{\mathrm{FLW}}      
            d\rho}}.\n{J2}
\end{equation}
We minimize \eq{J2}, finding the coefficients of $F_{1}=F_{\text{FLW}}$ and $F_{2}=F_{\text{our}}$. For the FLW numerical data $t_{\mu\nu}^{\mathrm{FLW}}(x)$, $J_{1}$ was minimized  at $J_{\text{FLW}}\approx 0.0000620$ with
\be
F_{\mathrm{FLW}}\! \approx\! 1\! -\! 1.062 \left(\frac{2M}{\rho}\right)
 \!+\! 0.554 \left(\frac{2M}{\rho}\right)^2
 \!-\! 0.120 \left(\frac{2M}{\rho}\right)^3\!\!.\n{FFLW}
\ee
For our numerical data $t_{\mu\nu}^{\mathrm{our}}(x)$,  the normalized mean-square error $J_2$ was minimized at $J_{\mathrm{our}} \approx 0.00139$ for
\begin{equation}
F_{\mathrm{our}}\! \approx\! 1\! -\! 1.002\left(\frac{2M}{\rho}\right)
 \!+\! 0.434 \left(\frac{2M}{\rho}\right)^2 
 \!-\! 0.059 \left(\frac{2M}{\rho}\right)^3\!\!.\n{FFour}
\end{equation}

Our data is less accurate than the FLW data, giving $J_{\mathrm{our}}\approx 22J_{\mathrm{FLW}}$. This is not unexpected, since we varied only 210 parameters in our spectral method, whereas FLW used grids of $40\times 40$ (or 1\,600 points) and of $160\times 160$ (or 25\,600 points).  Also, the individual coefficients of these two cubics have large relative differences, but the ratio of the two cubics themselves never differs by more than 1.3\% from unity, so they show good agreement between what is generated by our numerical data and by what is given by the FLW data. 

We then found an 11th-order polynomial $\text{F}_{11}$ to
the FLW data, which gave normalized mean-square error
$J_{11} \approx 0.0000572$ for
\ba
F_{11}&=&1-1.1241\left(\frac{2M}{\rho}\right)
 +1.956\left(\frac{2M}{\rho}\right)
 ^{2}-9.961\left(\frac{2M}{\rho}\right)
 ^3+35.475\left(\frac{2M}{\rho}\right)^4\nonumber\\
 &&-75.962\left(\frac{2M}{\rho}\right)^5
+99.432\left(\frac{2M}{\rho}\right)
 ^{6}
-73.694\left(\frac{2M}{\rho}\right)
 ^7+18.726\left(\frac{2M}{\rho}\right)
 ^8\nonumber\\
 &&+13.990\left(\frac{2M}{\rho}\right)^{9}
 -12.366\left(\frac{2M}{\rho}\right)
 ^{10}+2.900\left(\frac{2M}{\rho}\right)
 ^{11}.\n{FF11}
\ea

\begin{figure}[htb]
\centering
\subfigure[]{\includegraphics[width=8cm]{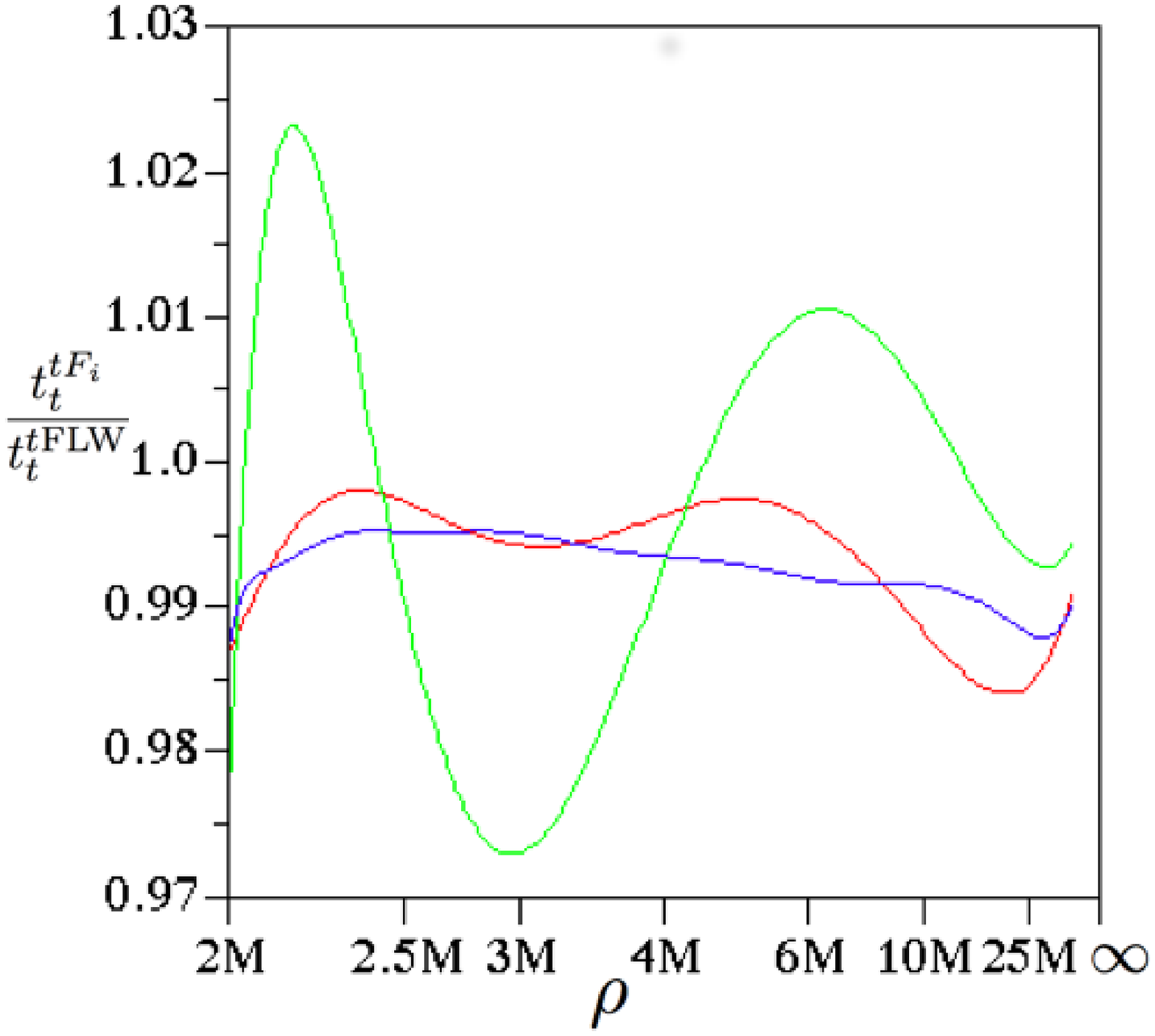}}
\subfigure[]{\includegraphics[width=7.3cm]{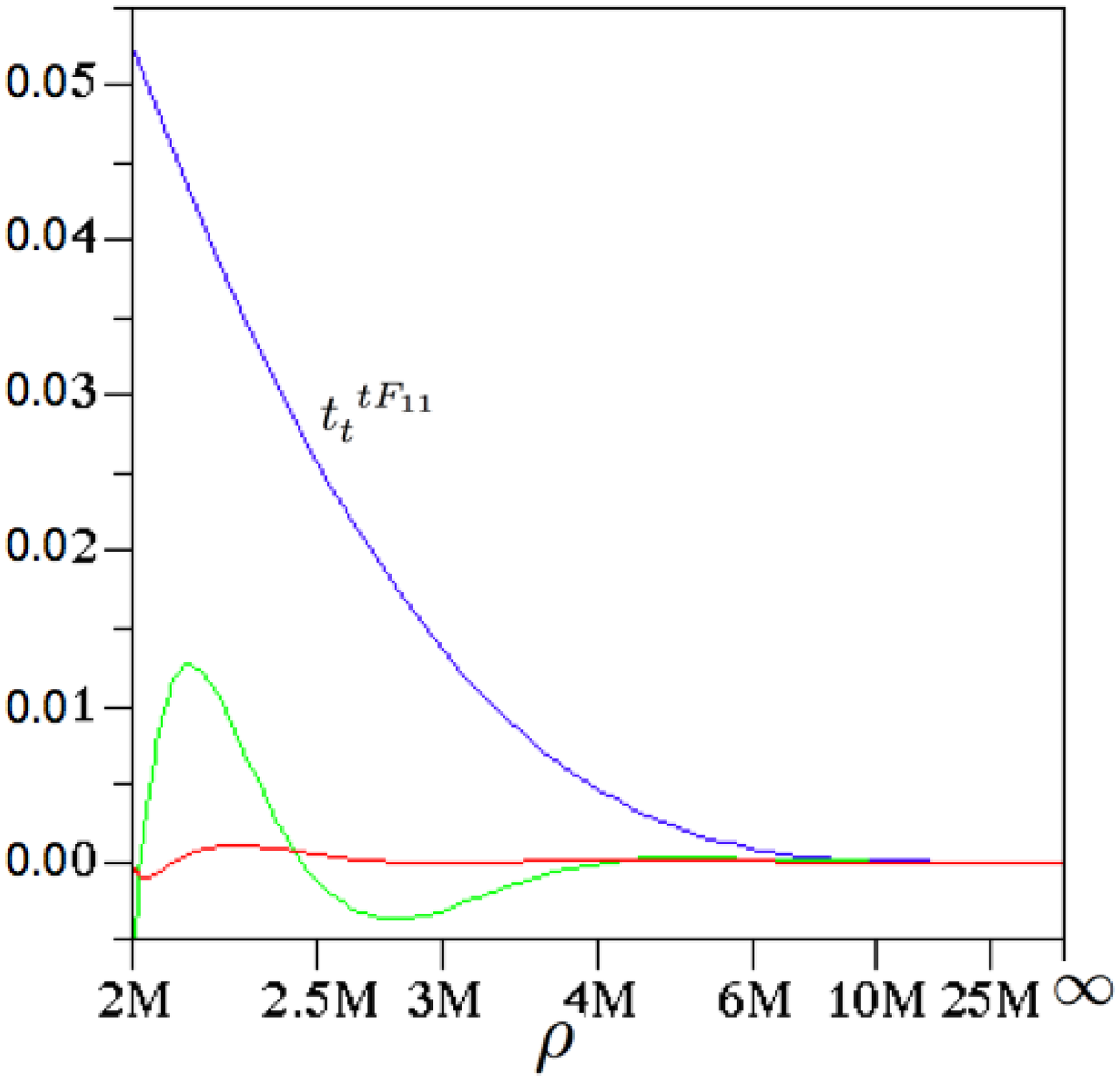}}
\caption{(a) The blue line is the ratio of the $t_{t}^{~tF_{11}}$, derived from the 11th-order polynomial fit $F_{11}$ to the FLW data, to $t_{t}^{~t\mathrm{FLW}}$, vs. $\rho$. The red line is the ratio of the $t_{t}^{~tF_{\mathrm{FLW}}}$, derived from the cubic fit $F_{\mathrm{FLW}}$ to the FLW data, to $t_{t}^{~t\mathrm{FLW}}$. The green line is the ratio of the $t_{t}^{~tF_{\mathrm{our}}}$, derived from the cubic fit $F_{\mathrm{our}}$ to our data, to $t_{t}^{~t\mathrm{FLW}}$. (b) The blue line is the $t_{t}^{~tF_{11}}$. The red line is the difference between the $t_{t}^{~tF_{\mathrm{FLW}}}$, derived from the cubic fit $F_{\mathrm{FLW}}$ to the FLW data, and the $t_{t}^{~tF_{11}}$, derived from the 11th order fit to the FLW data, times $10$, i.e., $10(t_{t}^{~t\mathrm{FLW}}-t_{t}^{~tF_{11}})$. The green line is the difference between the $t_{t}^{~tF_{\mathrm{our}}}$, derived from the cubic fit to our data, and the $t_{t}^{~tF_{11}}$, derived from the 11th order fit to the FLW data, multiplied by $10$, i.e., $10(t_{t}^{~t\mathrm{our}}-t_{t}^{~tF_{11}})$. }\label{F4.1} 
\end{figure}

\begin{figure}
\centering
\subfigure[]{\includegraphics[width=8cm]{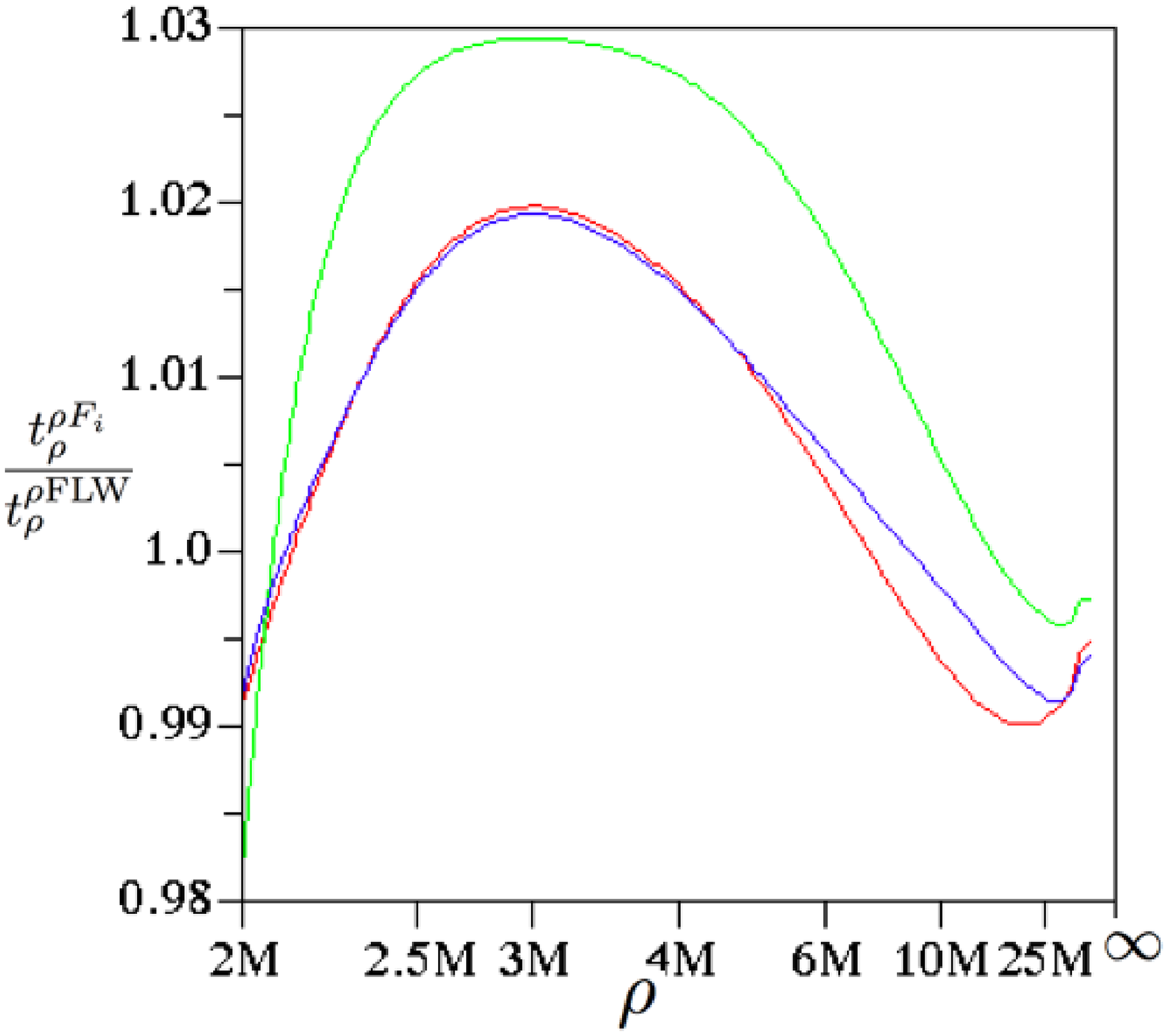}}
\subfigure[]{\includegraphics[width=7.3cm]{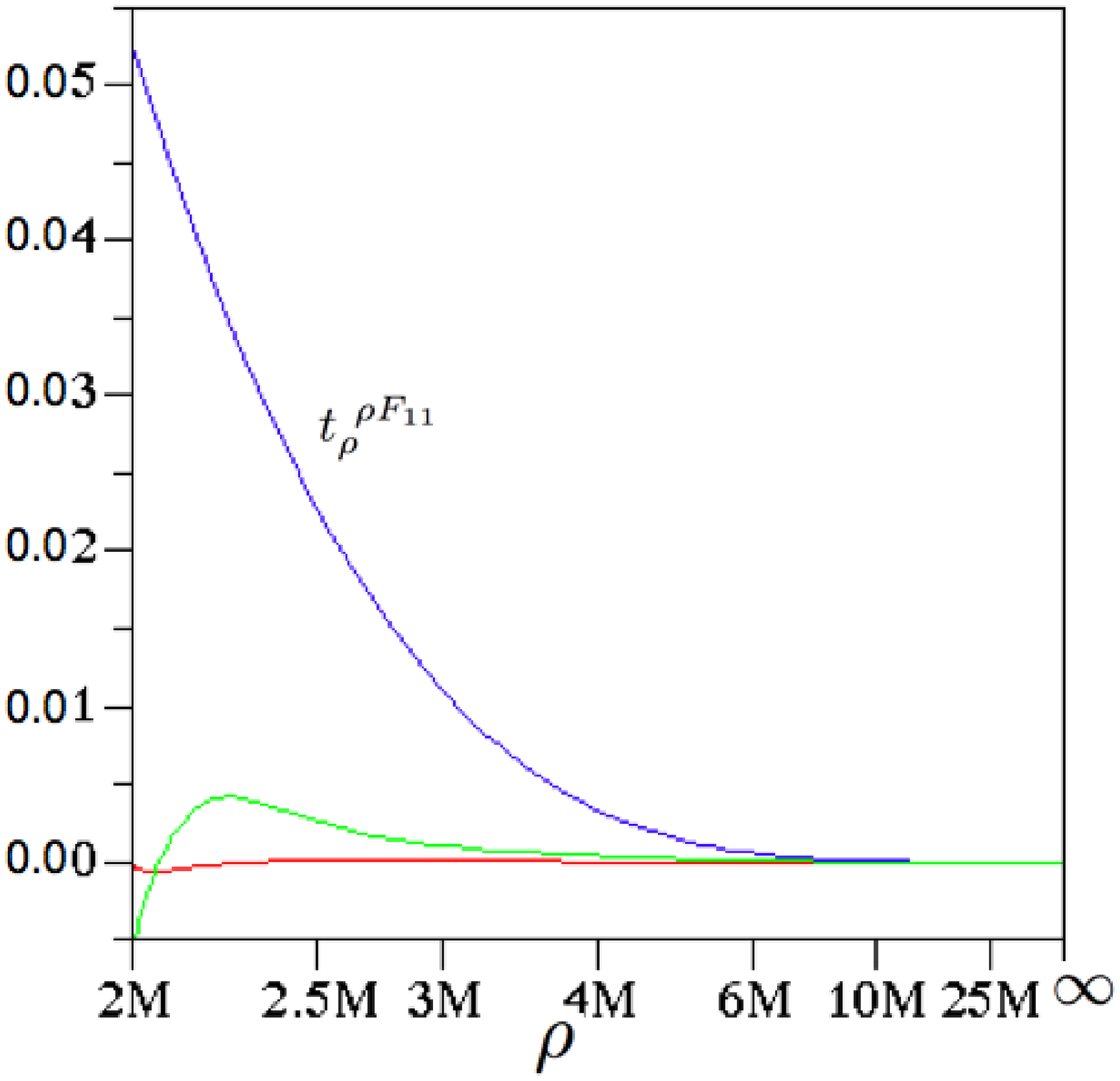}}
\caption{(a) The blue line is the ratio of the $t_{\rho}^{~\rho F_{11}}$, derived from the 11th-order polynomial fit $F_{11}$ to the FLW data, to $t_{\rho}^{~\rho F_{\mathrm{FLW}}}$. The red line is the ratio of the $t_{\rho}^{\rho F_{\mathrm{FLW}}}$, derived from the cubic fit $F_{\mathrm{FLW}}$ to the FLW data, to $t_{\rho}^{~\rho\mathrm{FLW}}$. The green line is the ratio of the $t_{\rho}^{~\rho F_{\mathrm{our}}}$, derived from the cubic fit $F_{\mathrm{our}}$ to our data, to $t_{\rho}^{~\rho\mathrm{FLW}}$. (b) The blue line is the $t_{\rho}^{~\rho F_{11}}$. The red line is 10 times the difference between the $t_{\rho}^{~\rho F_{\mathrm{FLW}}}$, derived from the cubic fit $F_{\mathrm{FLW}}$ to the FLW data, and the $t_{\rho}^{~\rho F_{11}}$, derived from the 11th order fit to FLW data, i.e., $10(t_{\rho}^{~\rho\mathrm{FLW}}-t_{~\rho}^{\rho F_{11}})$. The green line is $10(t_{\rho}^{~\rho F_{\mathrm{our}}}-t_{\rho}^{~\rho F_{11}})$. }\label{F4.2} 
\end{figure}

\begin{figure}
\centering
\subfigure[]{\includegraphics[width=8cm]{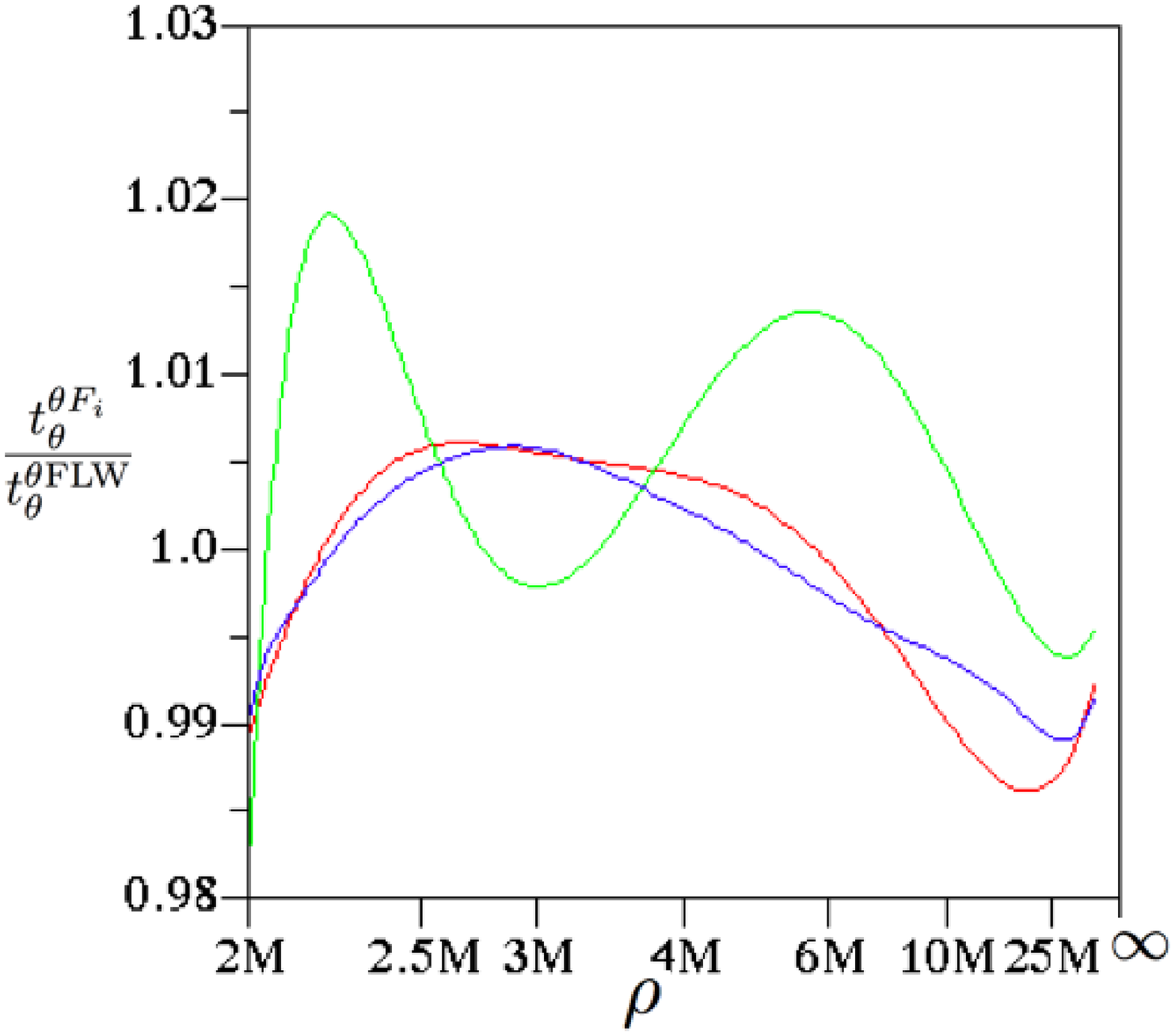}}
\subfigure[]{\includegraphics[width=7.3cm]{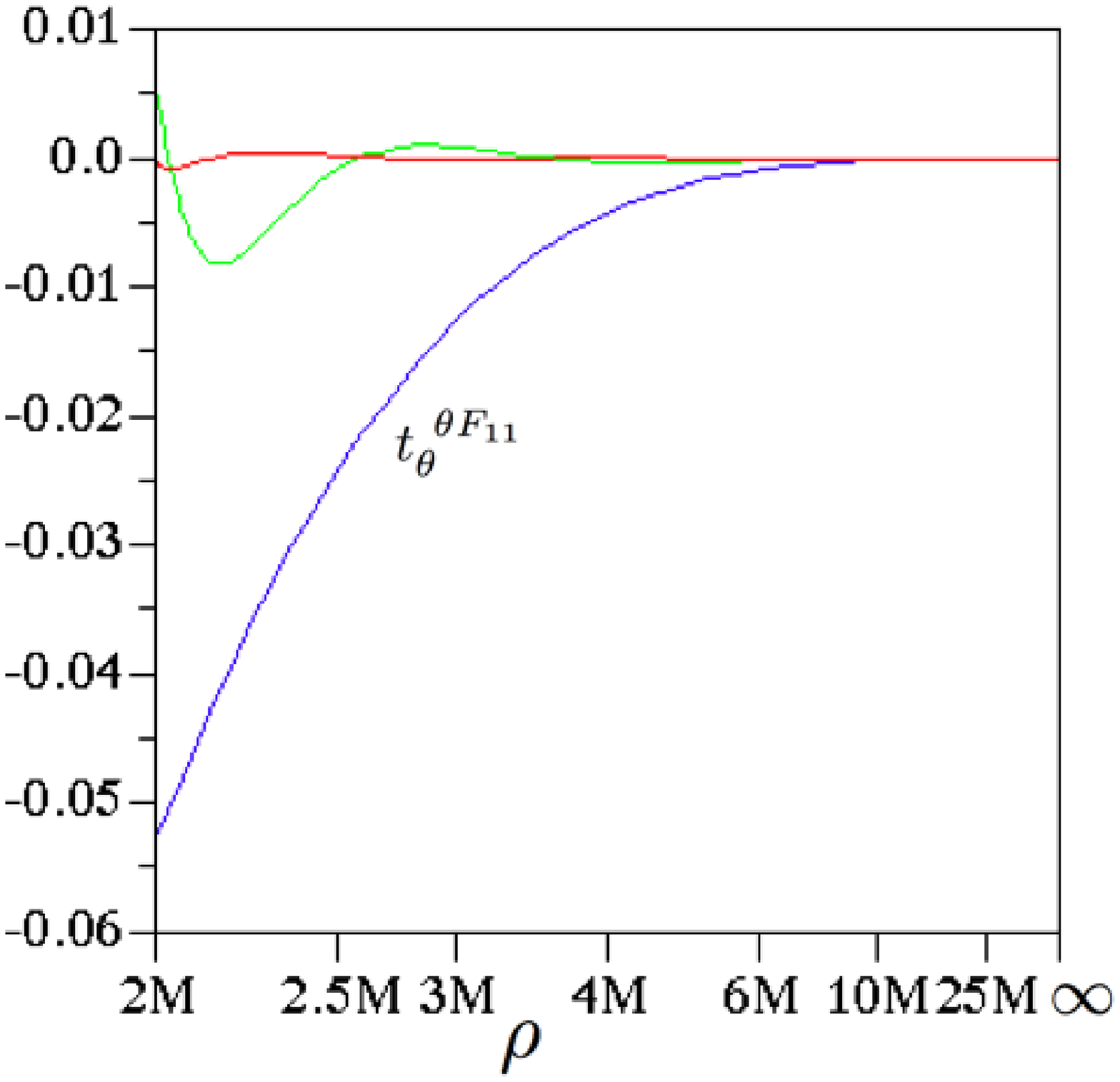}}
\caption{(a) The blue, red, and green lines are the ratio of $t_{\theta}^{~\theta F_{11}}$, $t_{\theta}^{~\theta F_{\mathrm{FLW}}}$, and $t_{\theta}^{~\theta F_{\mathrm{our}}}$ to the FLW data $t_{\theta}^{~\theta\mathrm{FLW}}$, respectively. (b) The blue line is the $t_{t}^{~tF_{11}}$. The red and green lines are $10$ times the difference of $t_{\theta}^{~\theta F_{\mathrm{FLW}}}$ and $t_{\theta}^{~\theta F_{\mathrm{our}}}$ with $t_{\theta}^{~\theta F_{11}}$, respectively.}\label{F4.3} 
\end{figure}

We have compared $F_{11}$, $F_{\mathrm{our}}$, and $F_{\mathrm{FLW}}$ in Fig \ref{F4.0}. Using $F_{\text{our}}$, $F_{\text{FLW}}$, and $F_{11}$, we can derive the corresponding $t_{\mu\nu}^{F_{11}}$ derived from the 11th order fit $F_{11}$ to the FLW data, $t_{\mu\nu}^{F_{\mathrm{FLW}}}$ derived from the cubic fit $F_{\mathrm{FLW}}$ to the FLW data, and $t_{\mu\nu}^{F_{\mathrm{our}}}$ derived from the cubic fit $F_{\mathrm{our}}$ to our data. In part (a) of Figs. \ref{F4.1}, \ref{F4.2}, and \ref{F4.3} we have considered the ratios of each of the three individual components of the three versions of $t_{\mu}^{\nu F_{i}}$ ($i=1$ for $F_{\mathrm{FLW}}$, $i=2$ for $F_{\mathrm{our}}$ and $i=11$ for $F_{11}$) to $t_{\mu}^{\nu FLW}$. These ratios were generally within $1-2\%$ of unity, with the maximum differing by less than $2.9\%$. 
\begin{table}
\begin{center}
\begin{tabular}{|c|c|c|c|c|}
\hline
~~~~~&$t_{\mu\nu}^{F_{11}}$ & $t_{\mu\nu}^{F_{\mathrm{FLW}}}$&$t_{\mu\nu}^{F_{\mathrm{our}}}$\\ \hline
$t_{\mu\nu}^{\mathrm{FLW}}$ &$J_{11}\approx 0.0000572$&$J_{\mathrm{FLW}}\approx 0.0000620$ &  
$J_{7}\approx 0.000214$\\ \hline
$t_{\mu\nu}^{F_{11}}$ & 0&$J_{5}\approx 0.000004793$ & $J_{6}\approx 0.000156$ \\ \hline\
$t_{\mu\nu}^{F_{\mathrm{FLW}}}$ &$J_{5}\approx 0.000004793$ &0& $J_{4}\approx 0.000146$ \\ \hline
\end{tabular}
\end{center}
\caption{Values of $J_{i}$ from Eq \eq{J}, with  $\Delta t_{\mu\nu}^{(i)}$ calculated as difference between the column $t_{\mu\nu}$ and the row $t_{\mu\nu}$.}
\label{Table2}
\end{table}

\begin{figure}[htb]
\begin{center}
\hspace{-1cm}
\subfigure[]{\includegraphics[width=7.7cm]{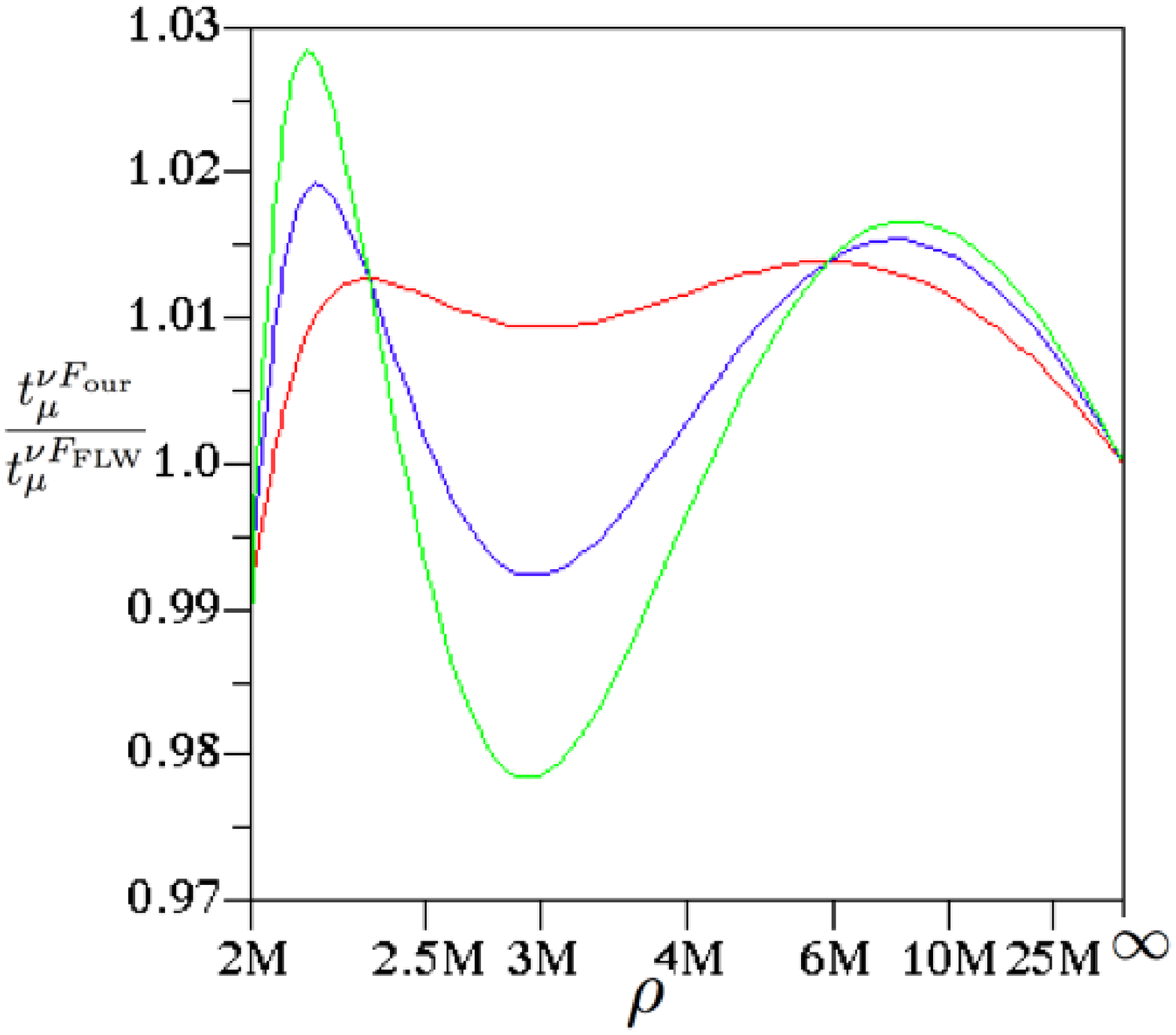}}
\subfigure[]{\includegraphics[width=7.7cm]{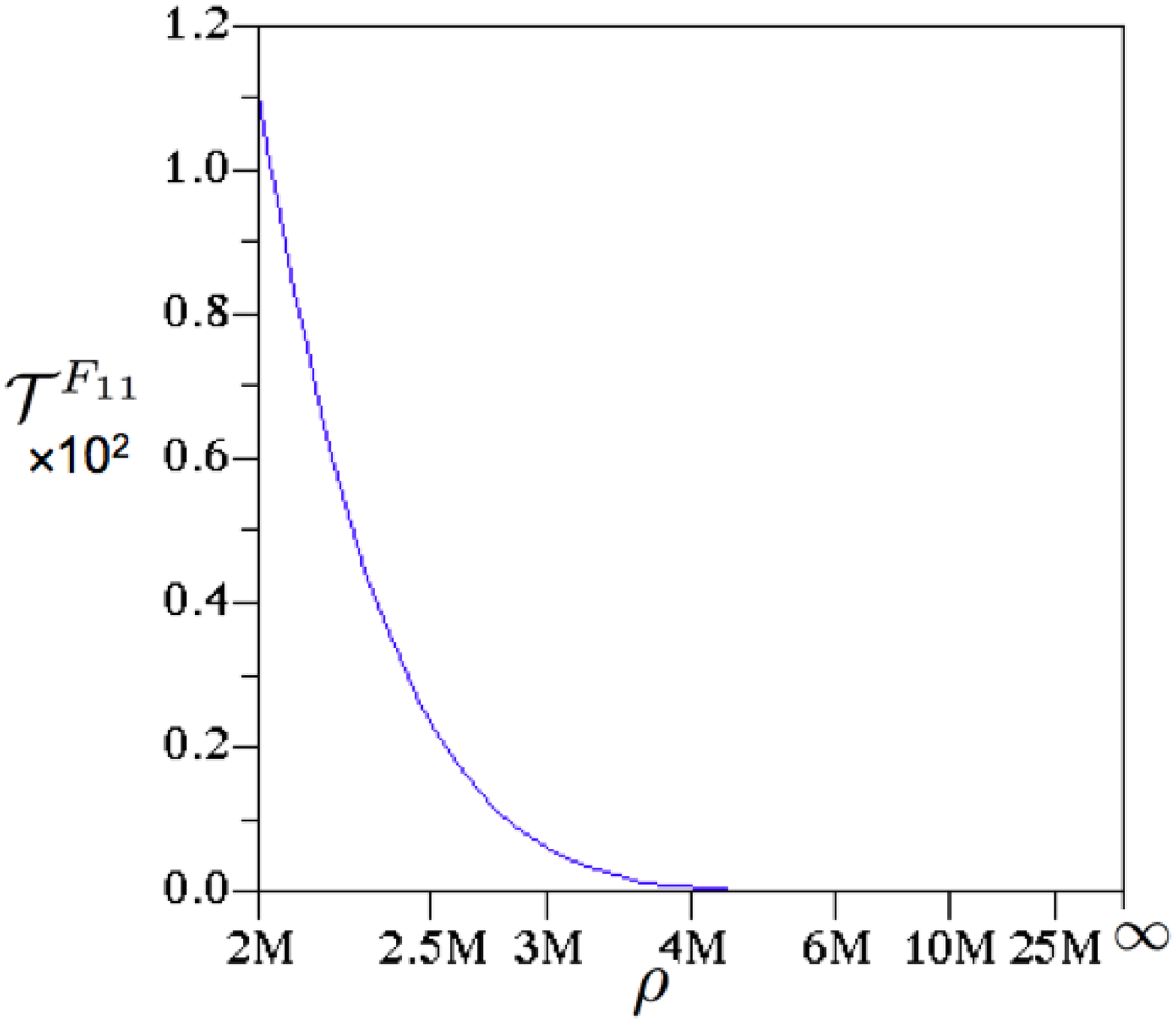}}
\caption{(a) The green, red, and blue lines are the ratios of $t_t^{~t F_{\mathrm{our}}}$, $t_\rho^{~\rho F_{\mathrm{our}}}$, and $t_\theta^{~\theta F_{\mathrm{our}}}$ to the corresponding $t_{i}^{~j F_{\mathrm{FLW}}}$. (b) $\mathcal{T}^{F_{11}}=t_{\mu\nu}^{F_{11}}t^{\mu\nu}_{F_{11}}$. }\label{F4.4} 
\end{center} 
\end{figure}

In part (b) of Figs. \ref{F4.1}, \ref{F4.2}, and \ref{F4.3} we have compared the difference between individual components of $t_{\mu}^{~\nu F_{\mathrm{our}}}$ with $t_{\mu}^{~\nu F_{11}}$ and $t_{\mu}^{~\nu F_{\mathrm{FLW}}}$ with $t_{\mu}^{~\nu F_{11}}$. The ratios of each of the three components of $t_{\mu\nu}^{F_{\mathrm{our}}}$ to $t_{\mu\nu}^{F_{\mathrm{FLW}}}$ as given in Fig \ref{F4.4} (a) is within $2.8\%$ of unity. 
The mean-square error between the $t_{\mu\nu}^{F_{\mathrm{our}}}(x)$ generated by our $F_{\mathrm{our}}$ fit to our data and the $t_{\mu\nu}^{F_{\mathrm{FLW}}}(x)$ generated by the $F_{\mathrm{FLW}}$ fit to the FLW data, using in Eq. (\ref{J2}) $\Delta t_{\mu\nu}^{(4)} = t_{\mu\nu}^{F_{\mathrm{our}}} - t_{\mu\nu}^{F_{\mathrm{FLW}}}(x)$, is $J_4 = J_{\mathrm{our\ fit\ vs.\ FLW\ fit}} \approx 0.000146 \approx 2.4\, J_{\mathrm{FLW}}$. Thus, the $t_{\mu\nu}^{F_{\mathrm{our}}}(x)$ generated by our $F_{\mathrm{our}}$ fits the $t_{\mu\nu}^{F_{\mathrm{FLW}}}(x)$ generated by the $F_{\mathrm{FLW}}$ fit to the FLW data nearly 9 times better than it fits the $t_{\mu\nu}^{\mathrm{our}}(x)$ directly extracted from our data, which is not quite traceless and conserved, as the $t_{\mu\nu}^{F_{\mathrm{our}}}(x)$ generated by the fitting $F_{\mathrm{our}}$ is constrained to be.

We also calculated $J_5 = J_{\mathrm{F_{11}\ fit\ vs.\ FLW\ fit}} \approx 0.000004793$, where now in Eq. \eq{J2} $\Delta t_{\mu\nu}^{(5)} = t_{\mu\nu}^{F_{11}} - t_{\mu\nu}^{F_{\mathrm{FLW}}}(x)$. Furthermore, we calculated $J_6 = J_{\mathrm{F_{11}\ fit\ vs.\ our\ F_{\mathrm{our}}\ fit}} \approx 0.000156$, where $\Delta t_{\mu\nu}^{(6)} = t_{\mu\nu}^{F_{11}} - t_{\mu\nu}^{F_{\mathrm{our}}}(x)$. The mean-square error between the $t_{\mu\nu}^{F_{\mathrm{our}}}(x)$ generated by our $F_{\mathrm{our}}$ fit to our data and the FLW data, $t_{\mu\nu}^{{FLW}}(x)$, using in Eq. (\ref{J2}) $\Delta t_{\mu\nu}^{(4)} = t_{\mu\nu}^{F_{\mathrm{our}}} - t_{\mu\nu}^{{FLW}}(x)$, is $J_7 = J_{\mathrm{our\ fit\ vs.\ FLW \ data}} \approx 0.000214$. For the summary of the comparison between different $t_{\mu\nu}$'s, refer to Table \ref{Table2}. 

\begin{figure}[htb]
\begin{center}
\hspace{-1cm}
\subfigure[]{\includegraphics[width=8cm]{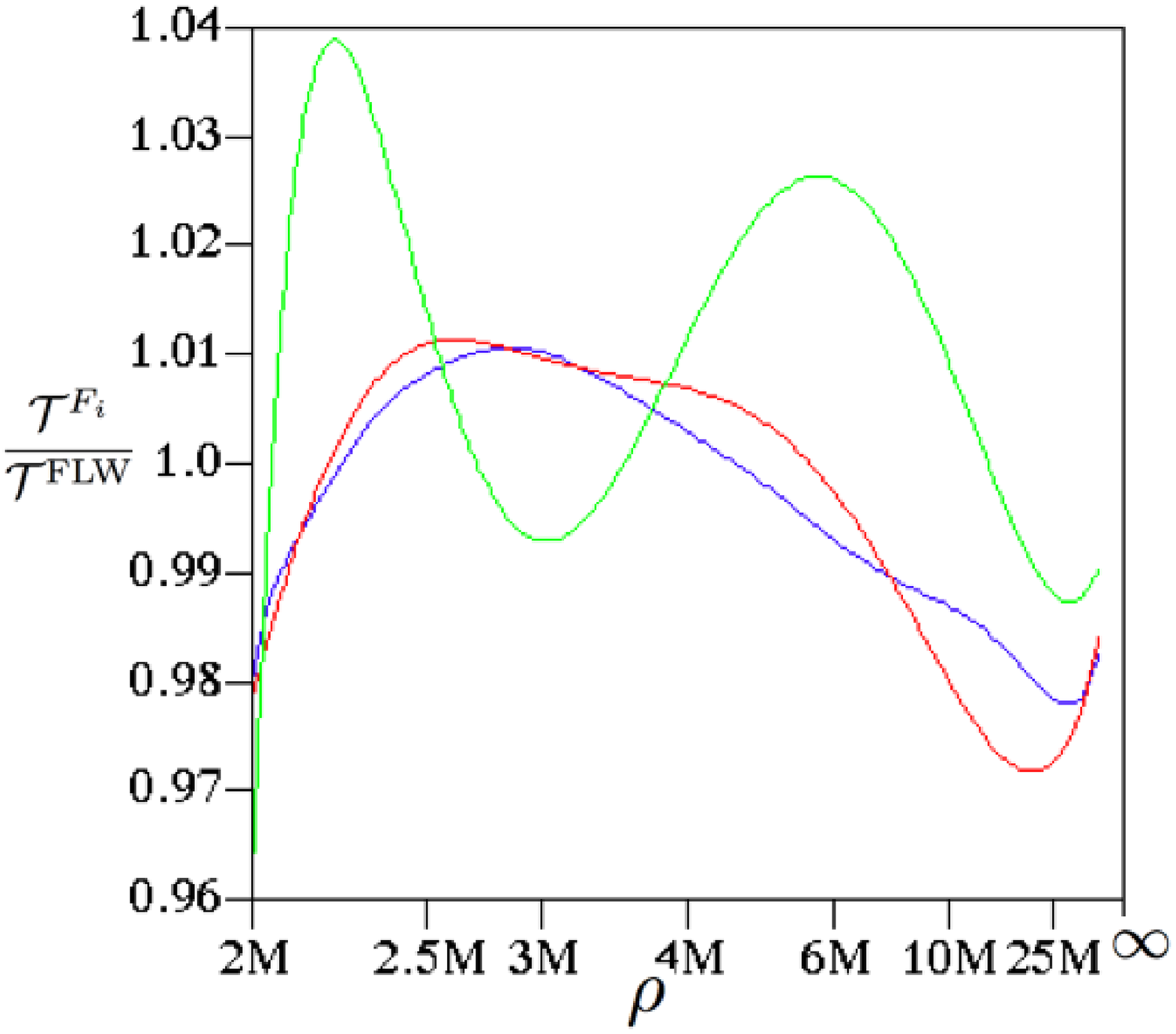}}
\subfigure[]{\includegraphics[width=7.8cm]{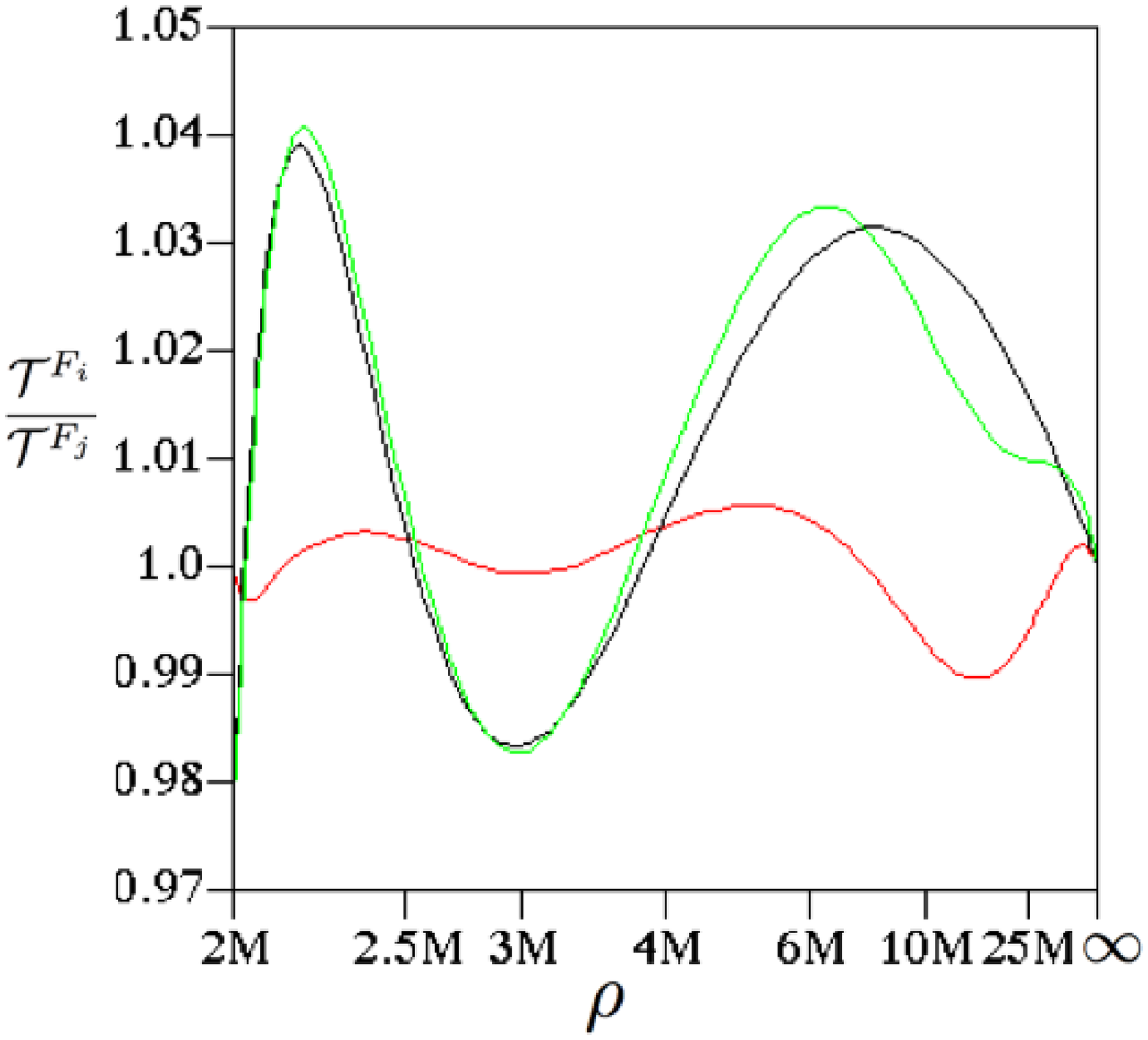}}
\caption{(a) The blue, red, and green lines are the ratios of $\mathcal{T}^{F_{\mathrm{11}}}$, $\mathcal{T}^{F_{\mathrm{FLW}}}$, and $\mathcal{T}^{F_{\mathrm{our}}}$ to $\mathcal{T}^{{\mathrm{FLW}}}$. (b) The violet, red, and green lines are the ratios $\mathcal{T}^{F_{\mathrm{our}}}/\mathcal{T}^{F_{\mathrm{FLW}}}$, $\mathcal{T}^{F_{\mathrm{our}}}/\mathcal{T}^{F_{11}}$, and $\mathcal{T}^{F_{11}}/\mathcal{T}^{F_{\mathrm{FLW}}}$, respectively. }\label{F4.5} 
\end{center} 
\end{figure}

Let us consider the trace of the square of energy momentum tensor, $\mathcal{T}=t^{\mu}_{\nu}t^{\nu}_{\mu}=(t_{t}^{~t})^{2}+(t_{\rho}^{~\rho})^{2}+2(t_{\phi}^{~\phi})^{2}$. We have plotted $\mathcal{T}^{F_{11}}$ derived from $t_{\mu}^{~\nu F_{11}}$'s in Fig \ref{F4.4} (b).  We made comparison between  $\mathcal{T}^{F_{i}}$'s and $\mathcal{T}^{FLW}$ which is directly given by the FLW data, Fig. \ref{F4.5} (a). The $\mathcal{T}^{F_{i}}$'s and $\mathcal{T}^{FLW}$ are in agreement within $3.9\%$. In addition from Fig \ref{F4.5} (b) the $\mathcal{T}^{F_{i}}$ and $\mathcal{T}^{F_{j}}$ are in agreement within $4\%$. 

\begin{figure}[htb]
\begin{center}
\hspace{-1.5cm}
\subfigure[]{\includegraphics[width=7.6cm]{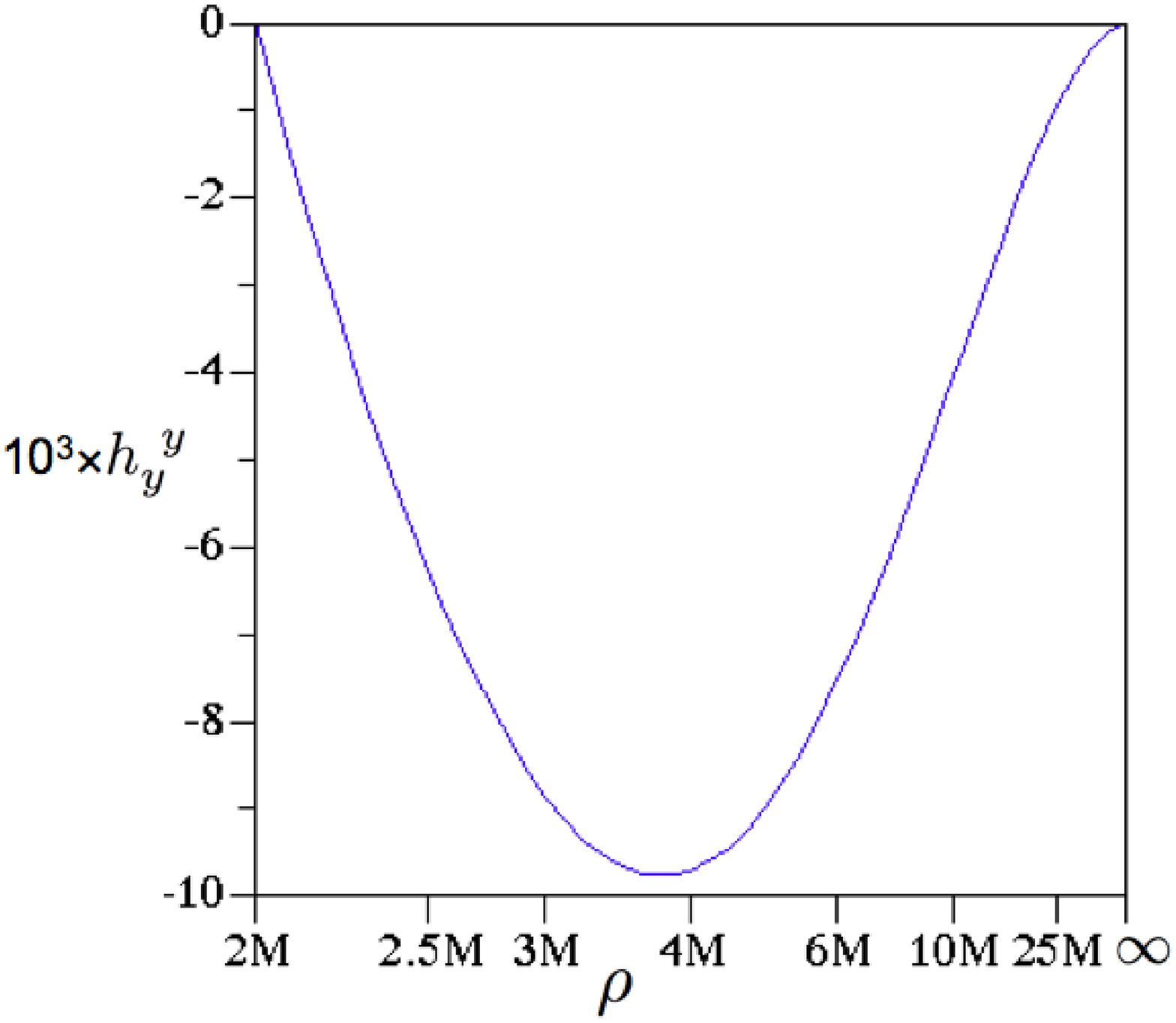}\label{F4.6a} }
\subfigure[]{\includegraphics[width=7.6cm]{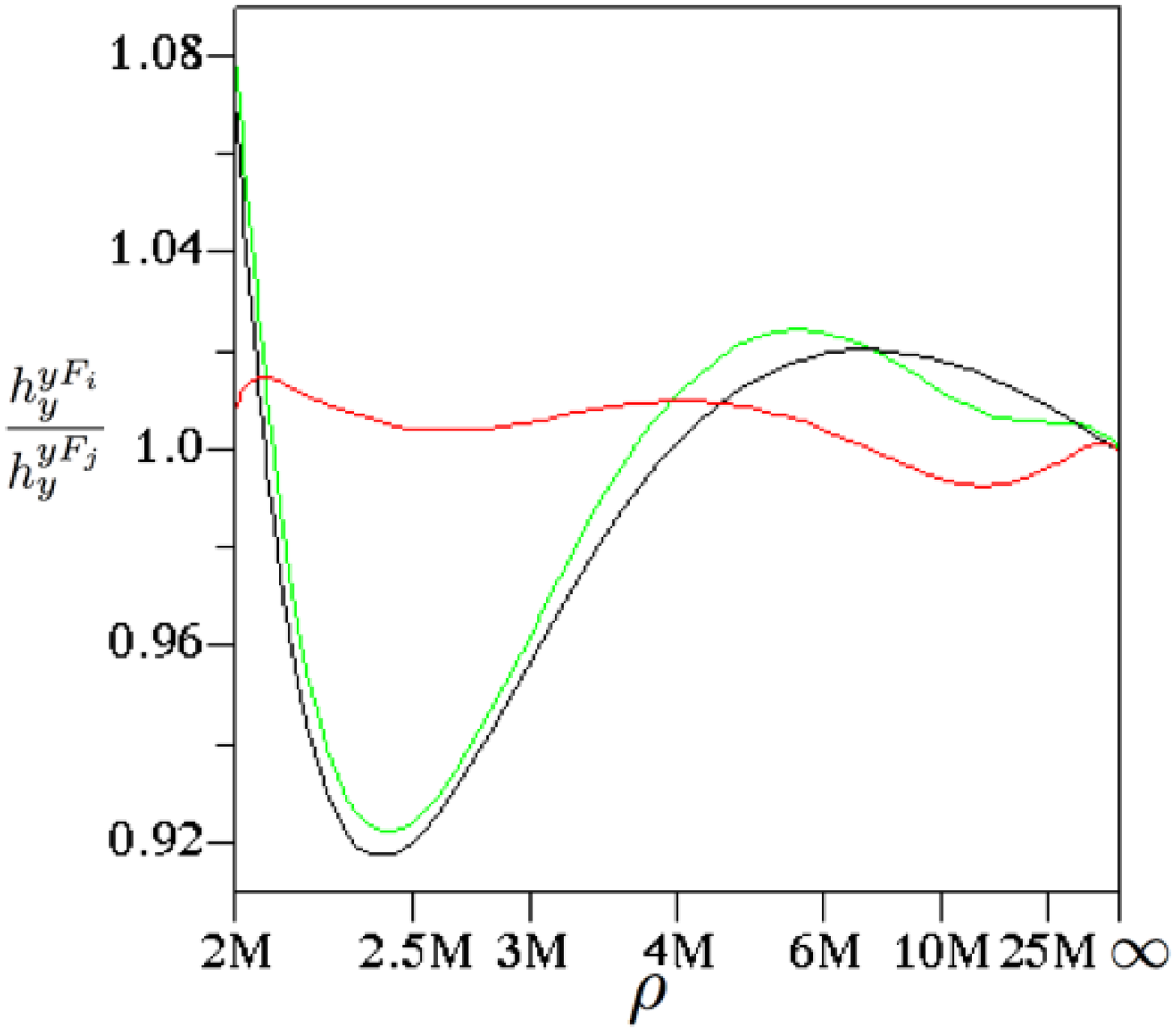}\label{F4.6b} }
\caption{(a) $h_{y}^{yF_{11}}$. (b) The violet, red, and green lines are the ratios ${h}_{y}^{~yF_{\mathrm{our}}}/h_{y}^{~yF_{\mathrm{FLW}}}$, $h_{y}^{~yF_{\mathrm{our}}}/h_{y}^{~yF_{11}}$, and $h_{y}^{~yF_{\mathrm{FLW}}}/h_{y}^{~yF_{11}}$, respectively.}\label{F4.6} 
\end{center} 
\end{figure}

\begin{figure}[htb]
\begin{center}
\hspace{-1.5cm}
\subfigure[]{\includegraphics[width=7.8cm]{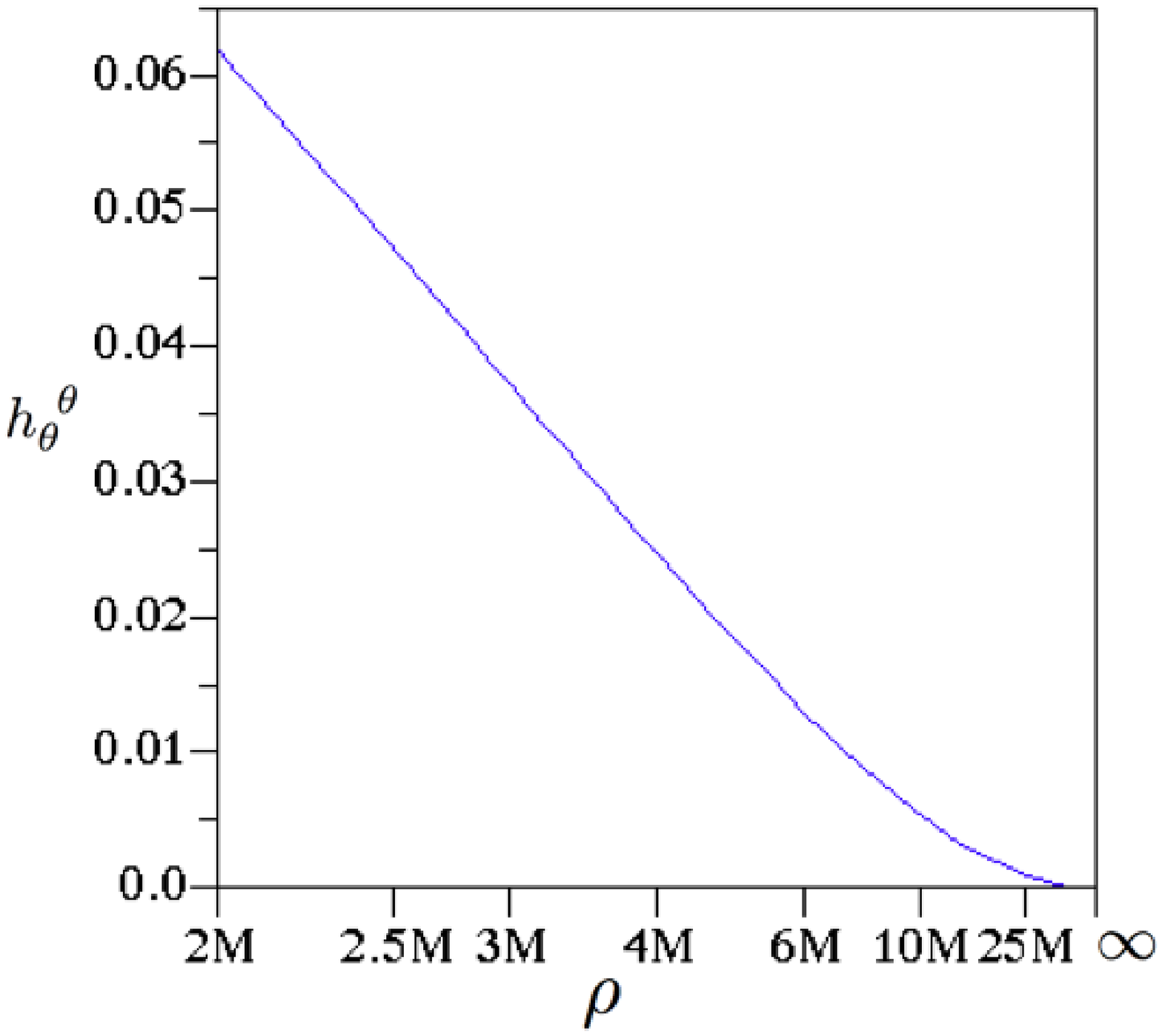}\label{F4.7a}}
\subfigure[]{\includegraphics[width=8cm]{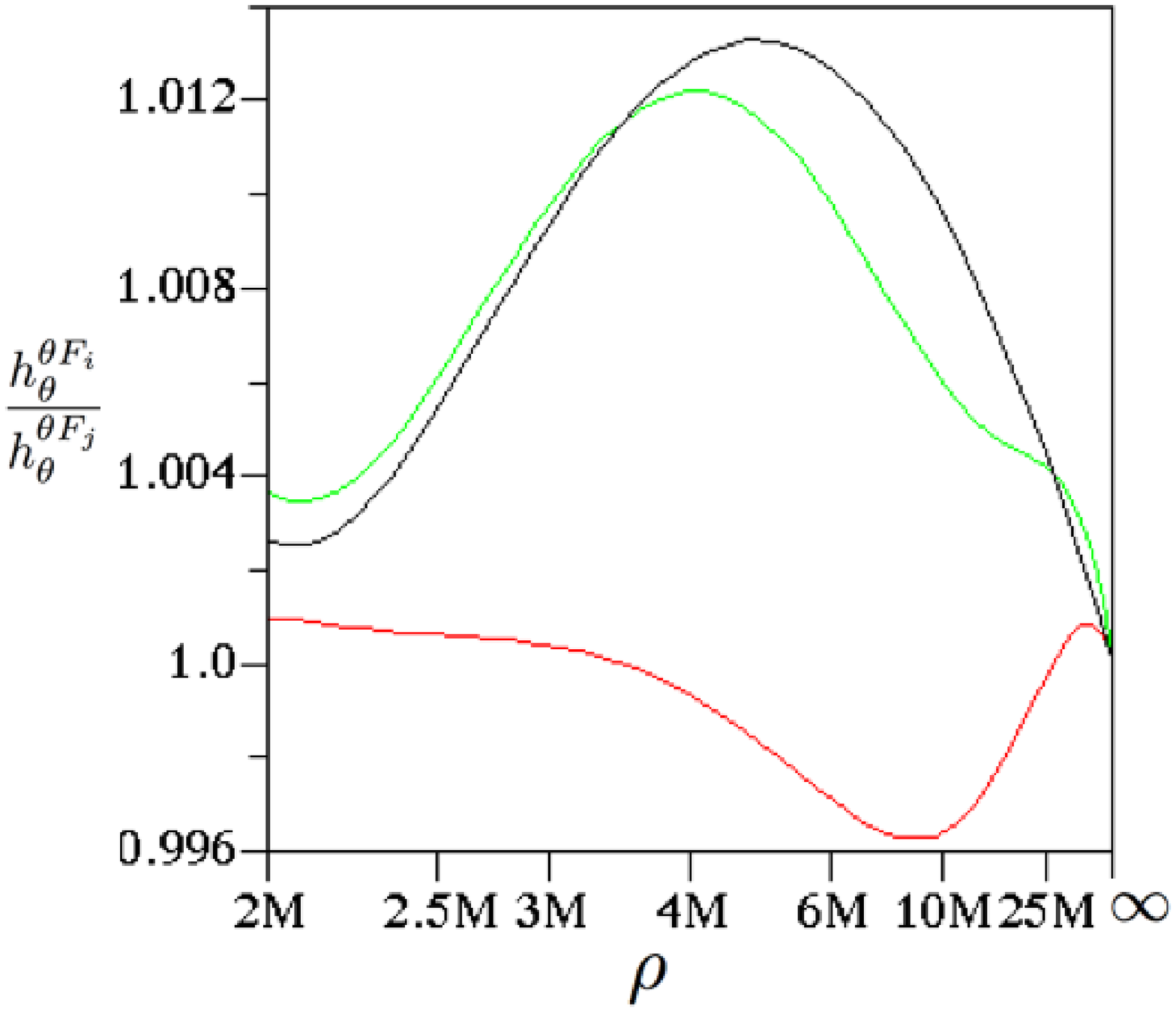}\label{F4.7b}}
\caption{(a) $h_{\theta}^{\theta F_{11}}$. (b)The violet, red, and green lines are the ratios ${h}_{\theta}^{~\theta F_{\mathrm{our}}}/h_{\theta}^{~\theta F_{\mathrm{FLW}}}$, $h_{\theta}^{~\theta F_{\mathrm{our}}}/h_{\theta}^{~\theta F_{11}}$, and $h_{\theta}^{~\theta F_{\mathrm{FLW}}}/h_{\theta}^{~\theta F_{11}}$, respectively. }\label{F4.7} 
\end{center} 
\end{figure}

We have calculated $h_{y}^{~y}$ from $F_{11}$, $F_{\mathrm{FLW}}$, and $F_{\mathrm{our}}$ using Eq. \eq{hyy}, where $h_{y}^{~y}$ generated from $F_{11}$ is shown in Fig. \ref{F4.6a}. The ratio of the $h_{y}^{~y}$ generated by $F_{\mathrm{our}}$ to $F_{\mathrm{FLW}}$, which involves a derivative of $F$ as given in Eq. \eq{hyy}, differ by up to about $9.3\%$, as shown in Fig. \ref{F4.6b}, while the ratio of the $h_{\theta}^{~\theta}=h_{\phi}^{~\phi}$, generated by using Eq. \eqref{F36}, $F_{\mathrm{our}}$, and $F_{\mathrm{FLW}}$ agree with unity to a very good precision of $1.3\%$, as shown in Fig. \ref{F4.7b}. 

\setcounter{topnumber}{2}
\setcounter{bottomnumber}{2}
\setcounter{totalnumber}{8}
\renewcommand{\topfraction}{0.85}
\renewcommand{\bottomfraction}{0.85}
\renewcommand{\textfraction}{0.15}
\renewcommand{\floatpagefraction}{0.7}

Having found $F$'s up to a good accuracy for both our numerical data and the FLW data, we can calculate some of the parameters of a RSII black hole on the brane as given by the metric \eq{MB}. The ADM mass, temperature, entropy and area of the RSII black hole on the brane, up to first order in $1/(-\Lambda M^{2})$, are 
\ba
 T &=&\frac{1}{8\pi M}+ O\left(\frac{1}{\Lambda^2 M^5}\ln{(-\Lambda M^2)}\right),\\
 S&=&4\pi M^2 + \mathrm{const} + O\left(\frac{1}{\Lambda^2 M^2}\ln{(-\Lambda M^2)}\right),\\
 A&=&16\pi M^2+ \frac{4\pi}{-\Lambda}F(1) + O\left(\frac{1}{\Lambda^2 M^2}\ln{(-\Lambda M^2)}\right).  
\ea 
\clearpage
For any $F$ that has only a constant term and negative powers of $\rho$, such as $F_{\mathrm{FLW}}$ given by Eq. \eq{FFLW}, $F_{\mathrm{our}}$ given by Eq. \eq{FFour}, and $F_{\mathrm{11}}$ given by Eq. \eq{FF11}, the ADM mass is precisely $M$ and the surface gravity of the black hole is $1/(4M)$.
 Aside from the numerical approximations for determining $F$, the metric \eq{MB} is only correct to first order in our perturbation parameter $1/(-\Lambda M^2)$, so there might be corrections to the surface gravity of a static RSII black hole to second order in $1/(-\Lambda M^2)$.  However, one can deduce that to first order in $1/(-\Lambda M^2)$, the Hawking temperature and entropy for the RSII black hole have the same values as they do for the Schwarzschild metric.

On the other hand, the horizon area is shifted from the Schwarzschild value $A_\mathrm{Sch} = 4\pi(2M)^2$ to $A_\mathrm{RSII} = 4\pi[(2M)^2 + F(1)/(-\Lambda)]$, where $F(1)$ is the value of $F$ on the horizon, at $y \equiv 2M/\rho = 1$. For the fit to the FLW data, $F_{\mathrm{FLW}}(1)\approx0.372$, which gives the area change $\Delta A=4\pi F(1)/(-\Lambda)\approx{4.67}/({-\Lambda})$, as compared to the Schwarzschild black hole with the same ADM mass $M$. For the fit to the FLW data with the 11th order polynomial $F_{\mathrm{11}}(1)$ is also $\approx 0.372$. For the fit to our numerical data, $F_{\mathrm{our}}(1)\approx 0.373$, or $\Delta A\approx{4.69}/({-\Lambda})$, which agrees within about $0.3 \%$ with FLW fit. 
Therefore the RSII black hole on the brane has a larger horizon area as compared to the Schwarzschild black hole with the same ADM mass $M$ by the amount $\Delta A = 4\pi F(1)/(-\Lambda) \approx 4.67/(-\Lambda)$, where here we used the FLW data value as probably more accurate. 
\vspace{-0.2cm}
\section{Conclusion}
We have constructed an infinite mass axisymmetric 5-dimensional black hole solution to the bulk Einstein equation with a cosmological constant $\Lambda=-6$. To do so, we have first expressed the metric \eq{m.1} in terms of the polynomial functions $A$, $B$, $C$, and $D$. We then have written these polynomial functions in terms of new ones $f$, $g$, $\tilde{A}$, $\tilde{B}$, $\tilde{C}$, and $\tilde{D}$, Eq. \eq{8m}, imposing the regularity conditions, and solving the Einstein equations to lowest order in $x$. Then, we have used a numerical approach, which to our knowledge is novel in the realm of general relativity, minimizing the integral of the squared error, Eq. \eq{1.8}, reducing it from $4038$ for $A=B=C=D=1$ to $4.2385\times 10^{-4}$ for 6th order polynomials with 210 coefficients. We have obtained a closed-form approximation
to the functions $A$, $B$, $C$, and $D$, and thus the metric \eq{m.1}.

We then have used this infinite mass 5-dimensional black hole solution to find the metric of a large Randall-Sundrum II black hole on the brane. Our result is independent numerical
evidence in support of the numerical discovery of
Figueras and Wiseman \cite{FW} of the existence of large static
black holes in the Randall-Sundrum II braneworld model \cite{A}, by a significantly
different numerical method. We have thoroughly compared our results to theirs and have shown that our
results agree quite well with theirs. We have obtained a good closed-form approximation to the metric of the black hole on the brane,
Eqs. \eq{MB} and either \eq{FFLW}, \eq{FFour}, or \eq{FF11}. Our
confirmation of the large black holes in RS II found by
Figueras and Wiseman \cite{FW}, and the fact that they are
very nearly the same as Schwarzschild black holes, show
that the RSII model can still be in agreement with astrophysical observations. If
large black holes did not exist in the Randall-Sundrum II braneworld model, the astrophysical observations of such black holes would have been strong evidence
against the viability of that model.

We have also shown the new result that to first order in our perturbation parameter
$1/(-\Lambda M^{2})$, the Hawking temperature and entropy
of the black hole is the same as that of a Schwarzschild
black hole of the same ADM mass M, while the RSII black hole on the brane has a larger horizon area as compared to the Schwarzschild black hole with the same ADM mass $M$ by the amount $\Delta A = 4\pi F(1)/(-\Lambda) \approx 4.67/(-\Lambda)$.
\newpage
\appendix
\section{Explicit Form of Functions}
The functions $A$, $B$, $C$ and $D$ in the metric \eq{m.1}, after numerically minimizing the integral of the squared error, Eq. \eq{1.8}, by 6th order polynomials with 210 coefficients, are given by the following expressions, with all of the coefficients rounded to five digits after the decimal place:
\begin{eqnarray}
A(x,y)&=&1+(0.21094+0.95771y-0.32215y^2+0.07863y^3\nonumber\\
&-&0.14553y^4+0.08707y^5-0.00708y^6)x\nonumber\\
&+&(0.483120-2.96522y+3.12221y^2+1.58758y^3\nonumber\\
&-&3.26594y^4+1.42017y^5-0.15995y^6)x^2\nonumber\\
&+&(-1.71060+15.96474y-30.20818y^2+13.67479y^3\nonumber\\
&+&7.52368y^4-6.45629y^5+0.54667y^6)x^3\nonumber\\
&+&(5.78445-61.41107y+170.77950y^2-214.55752y^3\nonumber\\
&+&153.70395y^4-73.81142y^5+20.14974y^6)x^4\nonumber\\
&+&(-11.34737+143.59002y-527.23697y^2+941.06600y^3\nonumber\\
&-&951.41715y^4+539.77303y^5-135.40474y^6)x^5\nonumber\\
&+&(14.60367-212.76740y+940.10896y^2-2014.17705y^3\nonumber\\
&+&2340.50264y^4-1423.20017y^5+356.46300y^6)x^6\nonumber\\
&+&(-11.53285+191.11234y-965.53690y^2+2306.98150y^3\nonumber\\
&-&2889.53569y^4+1827.70113y^5-460.15121y^6)x^7\nonumber\\
&+&(5.01719-94.50791y+529.64141y^2-1357.87635y^3\nonumber\\
&+&1776.26250y^4-1148.20207y^5+289.36969y^6)x^8\nonumber\\
&+&(-0.90738+19.65670y-120.15064y^2+323.38343y^3\nonumber\\
&-&433.62615y^4+282.63877y^5-70.80546y^6)x^9,\\
B(x,y)&=&1+(-0.60547-0.47885y+0.16108y^2-0.03931y^3\nonumber\\
&+&0.07277y^4-0.04354y^5+0.003540y^6)x\nonumber\\
&+&(0.43149-0.80874y+0.89420y^2-0.31160y^3\nonumber\\
&-&0.09760y^4+0.08441y^5-0.00482y^6)x^2\nonumber\\
&+&(0.21851+0.08687y-1.72664y^2+2.47145y^3\nonumber\\
&-&0.92513y^4-0.07910y^5+0.05733y^6)x^3\nonumber\\
&+&(-0.24942+0.72759y+3.13465y^2-9.10834y^3\nonumber\\
&+&8.03309y^4-3.11900y^5+0.42049y^6)x^4\nonumber\\
&+&(0.11567+0.85762y-8.90011y^2+22.33435y^3\nonumber\\
&-&24.96171y^4+13.30179y^5-2.72886y^6)x^5\nonumber\\
&+&(0.11933-2.07556y+11.88780y^2-30.32368y^3\nonumber\\
&+&37.81788y^4-22.84820y^5+5.46718y^6)x^6\nonumber\\
&+&(-0.03621+0.99966y-7.11882y^2+20.49026y^3\nonumber\\
&-&28.00052y^4+18.39565y^5-4.74288y^6)x^7\nonumber\\
&+&(0.00555-0.20539y+1.74472y^2-5.66200y^3\nonumber\\
&+&8.40417y^4-5.81216y^5+1.52634y^6) x^8,
\end{eqnarray}
\begin{eqnarray}
C(x,y)&=&1+(-0.60547-0.47885y+0.16108y^2-0.03931y^3\nonumber\\
&+&0.07277y^4-0.04353y^5+0.00354y^6)x\nonumber\\
&+&(-0.14516+1.32670y-1.67755y^2+0.52746y^3\nonumber\\
&+&0.82884y^4-1.10767y^5+0.49529y^6-0.06057y^7)x^2\nonumber\\
&+&(0.27707-2.97720y+6.80878y^2-6.66724y^3\nonumber\\
&+&2.34156y^4+2.29259y^5-2.95084y^6+0.978589y^7)x^3\nonumber\\
&+&(-0.54534+5.46359y-13.40233y^2+12.20018y^3\nonumber\\
&+&3.69101y^4-21.56518y^5+20.75440y^6-6.75727y^7)x^4\nonumber\\
&+&(0.63505-6.58020y+16.26668y^2-7.88244y^3\nonumber\\
&-&32.05193y^4+69.62487y^5-57.58350y^6+17.59022y^7)x^5\nonumber\\
&+&(-0.42810+5.02395y-14.87309y^2+10.13469y^3\nonumber\\
&+&30.26574y^4-73.46057y^5+62.70428y^6-19.32214y^7)x^6\nonumber\\
&+&(0.15406-2.23357y+9.56540y^2-15.94279y^3\nonumber\\
&+&4.13362y^4+20.02335y^5-24.13180y^6+8.41885y^7)x^7\nonumber\\
&+&(-0.01891+0.44363y-2.86261y^2+7.61814y^3\nonumber\\
&-&9.31272y^4+4.24741y^5+0.75762y^6-0.87134y^7)x^8,\\
D(x,y)&=&1+(0.39453-0.47885y+0.16108y^2-0.03931y^3\nonumber\\
&+&0.07276y^4-0.04354y^5+0.00354y^6)x\nonumber\\
&+&(0.05130+0.60697y-0.17110y^2-1.10809y^3\nonumber\\
&+&1.02899y^4-0.23688y^5-0.02783y^6)x^2\nonumber\\
&+&(0.32335-2.49967y+5.37277y^2-6.15484y^3\nonumber\\
&+&7.81122y^4-7.43134y^5+2.80293y^6)x^3\nonumber\\
&+&(-0.52530+7.18773y-31.23730y^2+79.39889y^3\nonumber\\
&-&119.57662y^4+94.40212y^5-29.80903y^6)x^4\nonumber\\
&+&(0.58030-12.77895y+87.87779y^2-284.24264y^3\nonumber\\
&+&463.83801y^4-369.81230y^5+115.08588y^6)x^5\nonumber\\
&+&(-0.13189+14.60642y-144.69672y^2+529.23363y^3\nonumber\\
&-&899.73602y^4+725.67077y^5-225.66361y^6)x^6\nonumber\\
&+&(-0.30734-11.10747y+144.89421y^2-565.76399y^3\nonumber\\
&+&982.58529y^4-797.70747y^5+248.00979y^6)x^7\nonumber\\
&+&(0.29947+5.31976y-82.06726y^2+330.24011y^3\nonumber\\
&-&579.34304y^4+472.44205y^5-147.08568y^6)x^8\nonumber\\
&+&(-0.08325-1.22603y+20.06377y^2-81.40275y^3\nonumber\\
&+&143.32172y^4-117.33319y^5+36.68467y^6)x^9. 
\end{eqnarray}
Note that after plugging the 6th order polynomials into Eq. \eq{8m}, the polynomials $A$, $B$, $C$, and $D$ end up having slightly higher order and $249$ coefficients, with $39$ of them not independent.
\acknowledgments
We have benefited from conversations with Pau Figueras, James Lucietti, and Toby Wiseman and greatly appreciate their sharing their detailed numerical data with us for comparison.  CC acknowledges an Avadh Bhatia Postdoctoral Fellowship at the University of Alberta.  This research was also supported in part by the Natural Sciences and Engineering Research Council of Canada.

\end{document}